\documentclass[apj,twocolumn]{emulateapj}
\usepackage{graphicx}
\usepackage{subfigure}
\usepackage{amsmath}
\usepackage{amsfonts}
\usepackage{amssymb}
\usepackage{xcolor}
\newcommand{\mpch}{\>h^{-1}{\rm {Mpc}}}

\begin{document}
\title{Two dimensional topology of cosmological reionization}

\author{Yougang Wang\altaffilmark{1}, Changbom Park\altaffilmark{2}, Yidong Xu\altaffilmark{1}, 
Xuelei Chen\altaffilmark{1,3}, Juhan Kim\altaffilmark{4}}

\altaffiltext{1}{Key Laboratory of Computational Astrophysics, National Astronomical Observatories, 
Chinese Academy of Sciences, Beijing, 100012 China; E-mail: wangyg@bao.ac.cn}

\altaffiltext{2}{School of Physics, Korea Institute for Advanced Study,85 Hoegiro, Dongdaemun-gu,
Seoul 130-722, Korea; cbp@kias.re.kr}

\altaffiltext{3}{Center for High Energy Physics, Peking University, Beijing 100871, China}

\altaffiltext{4}{Center for Advanced Computation, Korea Institute for Advanced Study, 85 Hoegiro, Dongdaemun-gu, Seoul 130-722, Korea}


\begin{abstract}

We study the  two-dimensional topology of  the 21-cm differential brightness temperature
for two hydrodynamic radiative transfer simulations and two semi-numerical models.
In each model, we calculate the two dimensional genus curve for the early, middle and 
late epochs of reionization.
It is found that the genus curve depends strongly on the ionized 
fraction of hydrogen in each model.  The genus curves are significantly different 
for different reionization scenarios even when the ionized faction is 
the same.  We find that the two-dimensional topology analysis method 
is a useful tool to constrain the reionization models.   
Our method can be applied to the future observations such as 
those of the Square Kilometer Array.

\end{abstract}


\maketitle

\section{Introduction}

Reionization is a milestone event in the history of the universe. 
Quasar absorption line observations indicate that its completion 
is at redshift $z\gtrsim 6.5$
\citep[e.g.][]{2006ARA&A..44..415F}. The history of the epoch of 
reionization (EoR) remains unclear, but the
cosmic microwave background (CMB) signal show that the total optical depth of
free electrons is $\tau=0.066 \pm 0.016$, corresponding to $z_{re}=8.8_{-1.4}^{+1.7}$ if the 
reionization happens suddenly at a redshift $z_{re}$ \citep{2015arXiv150201589P}, so its
beginning must be earlier. 
The complex reionization process has been investigated by many theoretical works
\citep[e.g.][] {2004ApJ...613....1F,2007MNRAS.380L...6C,2007MNRAS.375..324Z,2009ApJ...704.1396X,2009MNRAS.398.2122Y,2012ApJ...747..127Y,2013MNRAS.428.2467K}.
One of the most promising methods to detect the cosmic reionization
is through the 21 cm transition of HI. The emission or absorption of the 21 cm
line traces the neutral hydrogen well at different redshifts, which can
provide us with the most direct view of the reionization history.

Due to the strong foregrounds, the observation of
high redshift 21 cm signals is a challenging task. Although there
are a number of running radio interferometer arrays, such as the 21 Centimeter
Array \citep[21CMA;][]{2009AAS...21322605W}; the Giant Metre-wave Radio Telescope \citep[GMRT;][]{2013MNRAS.433..639P}, which gives a constraint on reionization at $z \approx8.6$;  the Low Frequency Array \citep [LOFAR;][]{2006astro.ph.10596R}, the Murchison Widefield Array \citep[MWA;][]{2013PASA...30...31B,2013PASA...30....7T};
and the Precision Array for Probing the Epoch of Reionization \citep[PAPER;][]{2015ApJ...801...51J},  which is designed for observing the
redshift 21cm signal from EOR and gives a new  $2\sigma$ upper limit on $\Delta^2(k)$ of ${\rm (22.4 mK)^2}$ in the range $0.15 < k < 0.5h{\rm  Mpc^{-1}}$ at z = 8.4.

Addtionally, 
the kinematic Sunyaev-Zel'dovich  \citep [kSZ;][]{1972CoASP...4..173S,1980MNRAS.190..413S} can distort the primary CMB black-body spectrum due to the peculiar velocity of the clusters of galaxies.  During the reionization, the ionized bubbles
generate angular anisotropy through the kSZ effect. The amplitude of  kSZ power depends on the process of reionization , and its shape depends on the distribution of bubble size. 
Therefore, the kSZ power spectrum can give constraints on the epoch of reioniztion \citep{2010PhRvD..81f7302M,2012ApJ...756...65Z,2012MNRAS.422.1403M,2015ApJ...799..177G}. When the post-reionization homogeneous kSZ signal is taken into account, \cite{2015ApJ...799..177G} found an upper limit on the duration $\Delta z<5.4$ at $95\%$ CL.

 At present, 
the reionization process is studied by 
numerical simulations \citep[e.g.][] {2006MNRAS.369.1625I,2007ApJ...671....1T,2008ApJ...689L..81T,2009ApJ...706L.164C,2012ApJ...747..127Y,2013ApJ...776...81B,2014MNRAS.439..725I}   or semi-numerical simulations \citep[e.g.][] {2007ApJ...669..663M,2011MNRAS.411..955M,2014MNRAS.443.2843M}, and different reionization 
scenarios can be explored by varying the input parameters of the models.  We expect the 21cm signal from the EoR to be detected in the near future, 
and the data will help constrain the theoretical models.

One popular way of analyzing 21 cm observation is to use the power spectrum of 
 the neutral hydrogen \citep{2007PhRvD..76h3005L,2011PhRvD..83j3006L,2012MNRAS.422..926M,2014ApJ...790..148M}. Recently, an
alternative approach based on the topological analysis has been used to quantify the
ionization status of the intergalactic medium   
\citep{2008ApJ...675....8L,2014JKAS...47...49H}. The topological
analysis was introduced to cosmology as a method to test the
Gaussianity of the primordial density field as predicted by many
inflationary scenarios \citep{1986ApJ...306..341G}, and this
tool has been developed and applied during the past 20 years
\citep{1986ApJ...309....1H,1987ApJ...319....1G,1989ApJ...340..625G,
2008ApJ...675...16G,2009ApJ...695L..45G,
1991ApJ...378..457P,1992ApJ...387....1P,2005ApJ...633....1P,2005ApJ...633...11P,
1994ApJ...420..525V}.  The genus of isodensity surfaces has been used to 
quantify the spatial distribution of galaxies and to constrain the galaxy
formation models  \citep{2005ApJ...633....1P,2010ApJS..190..181C}. 
Compared with the power spectrum, the topology method has certain niches. 
The topology of the isodensity contours 
is insensitive or less sensitive to nonlinear gravitational evolution, 
galaxy bias and the redshift distortion, since the the intrinsic 
topology does not change as the structures 
grow, at least not until the eventual break at 
shell crossing \citep{2005ApJ...633...11P,2010ApJ...715L.185P,2012ApJ...747...48W}.

In this paper, we use the 2d genus to characterize the different reionization models. 
Compared with the 3d genus, the 2d genus method can be applied to two-dimensional or nearly two-dimensional data sets, such as the cosmic microwave background. Although the three-dimensional data cube can be obtained in 
HI observations, in some cases the foreground removal process limit us to use the two-dimensional data. Secondly, it saves much time to compute the 2d genus than that the 3d one. One of the processes  (convolved by the smoothing filter) of the  genus calculation is the Fast Fourier Transform (FFT).
The speed of FFT is proportional to $NlogN$, where $N$ is the total of number of data points, which is related to the data points ($N_c$) in each dimension as $N=N_c^2$ and $N=N_c^3$ in two and three dimensions, respectively. Therefore,
the time taken  in 3d genus is at least  of $1.5N_c$ times in 2d genus.        
Even if the 3d genus can be obtained, the 2d genus can provide a useful cross check.


Below, we introduce the calculation of the 21 cm differential brightness temperature and
its 2d genus curve in Sec.2. In Sec. 3 we describe the radiative transfer simulations 
and semi-numerical simulation used in this paper. Our results are presented
Sec.4.  We summarize and discuss the results in Sec. 5.

\section{The two-dimensional genus of 21cm temperature}

\subsection{The 21 cm signal}
The emission or the absorption of 21 cm signal depends on the spin
temperature $T_{\rm S}$ of neutral hydrogen, which is defined  by the relative number
densities, $n_i$,  of atoms in the two
hyperfine levels of the electronic ground state,
$n_1/n_0 =3\exp(-T_{\star}/T_S),$ where $T_{\star}=h_p \nu_{10}/k_B=0.068\ K$  is the
equivalent temperature of the energy level hyperfine structure
splitting $h_p\nu_{10}=5.9\times10^{-6}$eV.   
The spin temperature $T_S$ is determined by several competing
processes (c.f. \citet{2006PhR...433..181F}):
\begin{equation}
T_S^{-1}=\frac{T_{\gamma}^{-1}+x_cT_K^{-1}+x_{\alpha}T_{C}^{-1}}{1+x_c+x_{\alpha}}
\end{equation}
where $T_{\gamma}=2.726(1+z)K$ is the cosmic microwave background (CMB) temperature at redshift
$z$, $T_K$ is the gas kinetic temperature, $T_C$ is the effective
color temperature of the UV radiation, $x_c$ is the coupling
coefficient for collisions and $x_{\alpha}$ is the coupling
coefficient for UV scattering. Comparing with the CMB temperature,
the 21 cm radiation is observed in emission if $T_S>T_{\gamma}$, or
absorption if $T_S<T_{\gamma}$. Generally, the 21 cm signal is
quantified by the 21 cm differential brightness temperature,
\begin{equation}\label{eq_tb1}
\delta T_b = \frac{T_S - T_{\gamma}}{1+z} \left( 1 - e^{-\tau}
\right) ,
\end{equation}
 where the optical depth $\tau$ is produced by a patch of neutral hydrogen, which
has the following form \citep{2006PhR...433..181F,2001PhR...349..125B}
\begin{eqnarray}\label{eq_tau}
\tau(z) &=& (2.8 \times 10^{-4}) \bigg(\frac{T_S}{1000\ {\rm K}}
\bigg)^{-1}\bigg(\frac{h}{0.70}\bigg)\bigg(\frac{1+z}{10} \bigg)^{3/2}  \nonumber \\
& & \times \bigg(\frac{\Omega_b}{0.046}
\bigg)\bigg(\frac{\Omega_m}{0.28} \bigg)^{-1/2}(1+\delta).
\end{eqnarray}
Here $\delta$ is the density contrast.
Assuming $\tau\ll 1$ and $T_s \gg T_{\gamma}$, then the brightness temperature of the
21cm emission can be written as
\begin{eqnarray}\label{eq_tb2}
\delta T_b &=& ({\rm 28\ mK})\bigg(\frac{h}{0.72}\bigg)\bigg(\frac{1+z}{10} 
\bigg)^{1/2}\bigg(\frac{H}{{\rm d}v_r/{\rm d}r+H}\bigg)  \nonumber \\
& & \times \bigg(\frac{\Omega_b}{0.044}
\bigg)\bigg(\frac{\Omega_m}{0.26} \bigg)^{-1/2}(1+\delta)(1-x_{i}).
\end{eqnarray}
where $x_{i}$ is the fraction of ionized hydrogen, $H(z)$ is the Hubble parameter, 
${\rm d}v_r/{\rm d}r$ is the comoving gradient of the line-of-sight (LOS) 
component of the comoving velocity. It is noted that the distribution
 of the brightness temperature of the neutral hydrogen  will be the same 
as that of  the underlying matter density field if
 ${\rm d}v_r/{\rm d}r$  is small (${\rm d}v_r/{\rm d}r<<H$) and the 
gas is fully neutral ($x_{\rm i}=0$).

\subsection{The 2d genus}
The 2d genus has been applied in many fields, such as the cosmic microwave background fluctuations \citep{1996MNRAS.281L..82C,1998ApJ...506..473P}, weak lensing field \cite{2001ApJ...552L..89M,2003ApJ...582L..67S}, non-Gaussian signatures \citep{2004MNRAS.349..313P}, large-scale structure in the redshift survey \citep{2007MNRAS.375..128J}, galaxy distribution in the Hubble deep fields \citep{2001ApJ...553...33P}, neutral hydrogen in both the Large and Small Magellanic Cloud \citep{2007ApJ...663..244K,2008ApJ...688.1021C}.     

If we consider the 21cm brightness temperature on a spherical shell, 
the 2d genus of the contour is defined by the number of contours
surrounding regions higher than a threshold value minus the number
of contours enclosing regions lower than the threshold
\citep{1989ApJ...345..618M,1990ApJ...352....1G,2013JKAS...46..125P}
\begin{equation}
G(\nu)=N_{\rm high}-N_{\rm low}
\end{equation}
where $N_{\rm high}$ and $N_{\rm low}$ are the number of isolated high-density regions
and low-density regions, respectively. The genus
$G(\nu)$  depends on the threshold density value $\nu$, which is in units of standard deviation from the mean.
Given a two-dimensional distribution, the 2d genus can be measured by using the  Gauss-Bonnett theorem \citep{1990ApJ...352....1G}
\begin{equation}
G_{2d}=\frac{\int C{\rm d}S}{2\pi}
\end{equation}
where the integral line is along the contour, and C is the inverse curvature $r^{-1}$ of the line. The value of  2d genus maybe negative or positive, depending on
whether a low-  or high-density region is enclosed.  If a curvature is integrated along a closed contour around a high-density region, its value will be 1, otherwise, its value
is -1.   For a Gaussian random field, the 2d genus per unit area  is given by \citep{1989ApJ...345..618M,1988MNRAS.234..509C}
\begin{equation}
G_{2d,Gaussian}=\frac{1}{(2\pi)^{3/2}}\frac{\langle
k^2\rangle}{2}\nu\exp(-\nu^2/2)
\end{equation}
where $\langle k^2\rangle=\int k^2P_2(k)d^2k/\int P_2(k)d^2k$ is the square of the wavenumber $k$ averaged 
over the smoothed two-dimensional  power spectrum $P_2(k)$. In practice, the one-point distribution of the density field is
not interesting \citep{1994ApJ...420..525V,2001ApJ...553...33P} , we follow \citep{2001ApJ...553...33P} to the parametrize the area fraction by
\begin{equation}
f_A=\frac{1}{\sqrt{2\pi}}\int_{\nu_A}^\infty e^{-t^2/2} {\rm d}t
\end{equation}   
The genus is calculated from $\nu_A=-3$ to 3 within an interval of 0.2. Here we use the
numerical code contour2d to calculate the 2d genus
\citep{1986ApJ...306..341G,1989ApJ...345..618M}.  In this code, the 2d genus is calculated by counting the turning of a 
contour observed at each vertex of four pixels: 1/4 contribution is from each vertex with 1 high density-region pixel and 3
low-density region pixels, -1/4 from each vertex with 3 high density-region pixel and 1
low-density region pixels, and otherwise is zero.

\section{Simulations}

Our reionization models are different
from what was used in the work of \cite{2014JKAS...47...49H}. They used the N-body
simulation and $\rm{C^2-Ray}$ \citep{2006NewA...11..374M} method. Here we
use the hydrodynamic radiative transfer (HRT) simulations \citep{2008ApJ...689L..81T}
 and the semi-numerical model 21cmFAST \cite{2011MNRAS.411..955M}.
The main advantage of the HRT simulation over the simulation 
in \cite{2014JKAS...47...49H} is that
 it  is a real hydrodynamical simulation which keeps track of baryon evolution,
heating, and cooling processes. The 21cmFAST routine is an approximation to 
the HRT simulation, and it is faster than both simulation in \cite{2014JKAS...47...49H}
and the HRT simulation.

Throughout the paper, we use the Wilkinson Microwave Anisotropy
Probe (WMAP) 5 year cosmological parameters: $\Omega_m=0.258$,
$\Omega_{\Lambda}=0.742$, $\Omega_b=0.044$, $h=0.719$,
$\sigma_8=0.796$, and $n_s=0.963$ \citep{2009ApJS..180..306D}, which are consistent with 
the cosmology parameters in the HRT simulation used in this study.

\subsection{cosmological radiative transfer simulation}
The hydrodynamic radiative transfer  simulations used in this
paper was described in detail in  \cite{2008ApJ...689L..81T}. The simulation is
based on the numerical method described in
\cite{2007ApJ...671....1T},  which includes an $N$-body algorithm for
dark matter, a star formation prescription, and a radiative transfer
algorithm for ionizing photons.  The $N$-body simulation includes $3072^3$ 
dark matter particles on an effective mesh with $11,520^3$ cells in a comoving  
box, $100 \mpch$ on each side.  The mass of
each dark matter particle is $2.68\times10^6M_{\odot}$. The resolution of hydrodynamic+RT simulations 
is $N=1536^3$ of dark
matter particles, gas cells, and adaptive rays.  
The photoionization and photoheating rates are calculated for each cell. 

Star formation occurs for particles with the 
density $\rho_m>100\rho_{\rm crit} (z)$ and temperature $T>10^4$ K.   
This cut in the temperature-density phase space restricts star formation effectively
to regions within the virial radius of halos that cool efficiently 
through atomic line transitions. 

Here we use two groups of the HRT simulations, which have different
finishing time for the hydrogen reionization. In the first simulation,
the reionization is completed  late at  $z\sim 6$ (HRT sim1),
while in the second simulation the reionization is finished early 
at  $z\sim 9$ (HRT sim2).

\subsection{The semi-numerical simulation 21cmFAST}
We also use a semi-analytical code 21cmFAST \citep{2011MNRAS.411..955M}
to study the cosmological 21 cm signal.  The  21cmFAST code  is a useful
semi-numerical code to model the reionization process.
Given the box size and particle number, the Gaussian random initial conditions of the 
dark matter density and velocity fields are generated by Monte Carlo 
sampling method. The large-scale density and peculiar velocity field
are then obtained by first-order perturbation theory.  
Assuming that the number of ionizing photons are 
proportional to the collapse fraction computed from the 
extended Press-Schechter formalism, the ionization field is generated from the
evolved density field at each redshift. From the density, 
ionization and peculiar velocity,  the 21cm brightness temperature is obtained.  
The advantage of this approach is that it 
is very fast to calculate the 21 cm signal for different model parameters. 

In order to match the completion time of the cosmic reionization in the
HRT simulation, we change the ionizing efficiency factor $\zeta$, 
which is defined as $\zeta=f_{\rm esc}f_{\star}N_{\gamma/b}n_{\rm rec}^{-1}$. 
Here $f_{\rm esc}$ is the escape fraction of ionizing photons from the object, 
$f_{\star}$ is the star formation efficiency, $N_{\gamma/b}$ is the number of 
ionizing photons produced per baryon in stars, and $n_{\rm rec}$ is the 
typical recombined number of times for a hydrogen atom. The box size 
is the same as in HRT simulations, the cell number is $N=768^3$. The cosmology 
parameters in our 21cmFAST simulation are chosen to be the same 
as in the HRT simulation, which is based on the
WMAP5 data. We have run two simulations ($\zeta=15$ for 
21cmFAST1 and $\zeta=50$ for 21cmFAST2) by
using the 21cmFAST code.  The finishing time of reionization  
is $z\sim6$ in 21cmFAST1 and $z\sim9$ in 21cmFAST2.
Combined with two HRT simulations, we define  HRT sim1 
and 21cmFAST1 as the two \emph{late models}, 
while HRT2 and 21cmFAST2 as the two \emph{early models}. 

\begin{figure}
\resizebox{\hsize}{!}{\includegraphics[width=\textwidth]{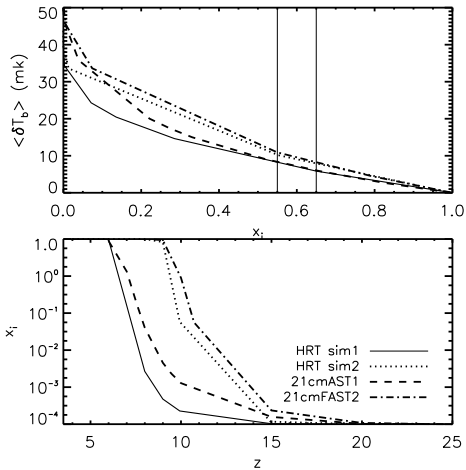}}
\caption{Lower Panel:Evolution of the ionization fraction as a
function of redshift for the different models. 
Upper Panel: Relation between the ionization fraction 
and the average differential brightness temperature of
21cm signal. The  two thin vertical lines indicate $x_{i}=0.55$ and $x_{i}=0.65$, 
respectively. } \label{fig:xHI}
\end{figure}

It is interesting to compare the different models at a fixed ionization
fraction.  Therefore, we output one snapshot of 21cmFAST1 simulation
which has the same ionization  fraction as the one from HRT sim1 
at $x_{i}=0.65 $, and another snapshot from 21cmFAST2 with
the same ioniztion fraction from HRT sim2 at $x_{i}=0.55 $. 
The two thin vertical lines in Figure~\ref{fig:xHI} show
$x_{i}=0.55$ and $x_{i}=0.65$.        

For the 21cmFAST simulation, we shall compare the 2d genus 
curve of the differential  brightness 
temperature with that of the matter distribution. For the HRT 
simulations, the dark matter and gas components are separate, 
so we shall show the genus curve of gas for comparison.





In Figure~\ref{fig:xHI}, we show the evolution of ionization fraction of  the neutral
hydrogen fraction (low) and the relation between the ionization
fraction and the average differential brightness
temperature (up) of 21 cm signal in four simulations. 
The ionization fraction $x_{i}$ increases 
rapidly with the decreasing redshift, 
and the mean differential brightness temperature decreases rapidly with time.    
In other words, the full reionization processes are fast in these models.  
the HRT sim1 and the 21cmFAST1 have the same reionization completion time
and the same ionization fraction at one redshift, but the ionization
fractions at other redshifts are different.  The simulations HRT
sim2 and 21cmFAST2 have nearly the same finishing time of the reionization. 
Compared with the two late models,  the difference of the ionization  
fraction evolution and the $x_{i}$-$\delta T_b$ relation 
in the two early models are  smaller. This can help us 
to discriminate the  different reionization scenarios even
if they have similar ionization fractions.

In real observations, the observed $\delta T_b$ includes both signal and noise. To study the effect of the 
thermal noise on the 2d genus, we add a Gaussian noise to the signal map. We 
generated 500 maps with the random noise following the Gaussian distribution, 
where the stand deviation is $f \sigma_T$. 
where $f$ is a constant, and 
$$\sigma_T=\sqrt{\frac{\delta T_b-<\delta T_b>}{N_c^2-1} }.$$

\section{results}
The observed signal can be characterized as the real signal convolved 
with the telescope response function or beam, however, the real  telescope beam is complicated. 
To model the observed signal, in this paper we assume either a Gaussian or compensated Gaussian 
lobe function for simplicity.  The Gaussian beam is simple, and widely used in the radio studies ~\citep[e.g.][]{2014ApJ...790..148M,2014MNRAS.441.3271W} to model the actual beam, which can be written as 
\begin{equation}
F_G(\theta)=\frac{1}{2\pi\sigma^2}\exp\bigg(-\frac{\theta^2}{2\sigma^2}\bigg).
\end{equation}
The full width at half maximum (FWHM) for the Gaussian beam is
$\Delta\theta=2\sigma\sqrt{2\ln2}$. Another popular choice is the
compensated  Gaussian,  
\begin{equation}\label{eq:fcg}
F_{CG}(\theta)=\frac{1}{2\pi\sigma^2}\bigg(1-\frac{\theta^2}{2\sigma^2}\bigg)\exp\bigg(1-\frac{\theta^2}{2\sigma^2 }\bigg).
\end{equation}   
The compensated Gaussian function approximates well the observational beam shape of a compact 
interfermeter array (often referred to as `dirty beam'), which is insensitive to large-scale features. 
Eq.~(\ref{eq:fcg}) shows that $F_G<0$ if $\theta>\sqrt{2}\sigma$, i.e. 
the sign of the contribution is negative. 
 
\begin{figure*}[p]
\subfigure[HRT sim 1]{\includegraphics[width=0.45\textwidth]{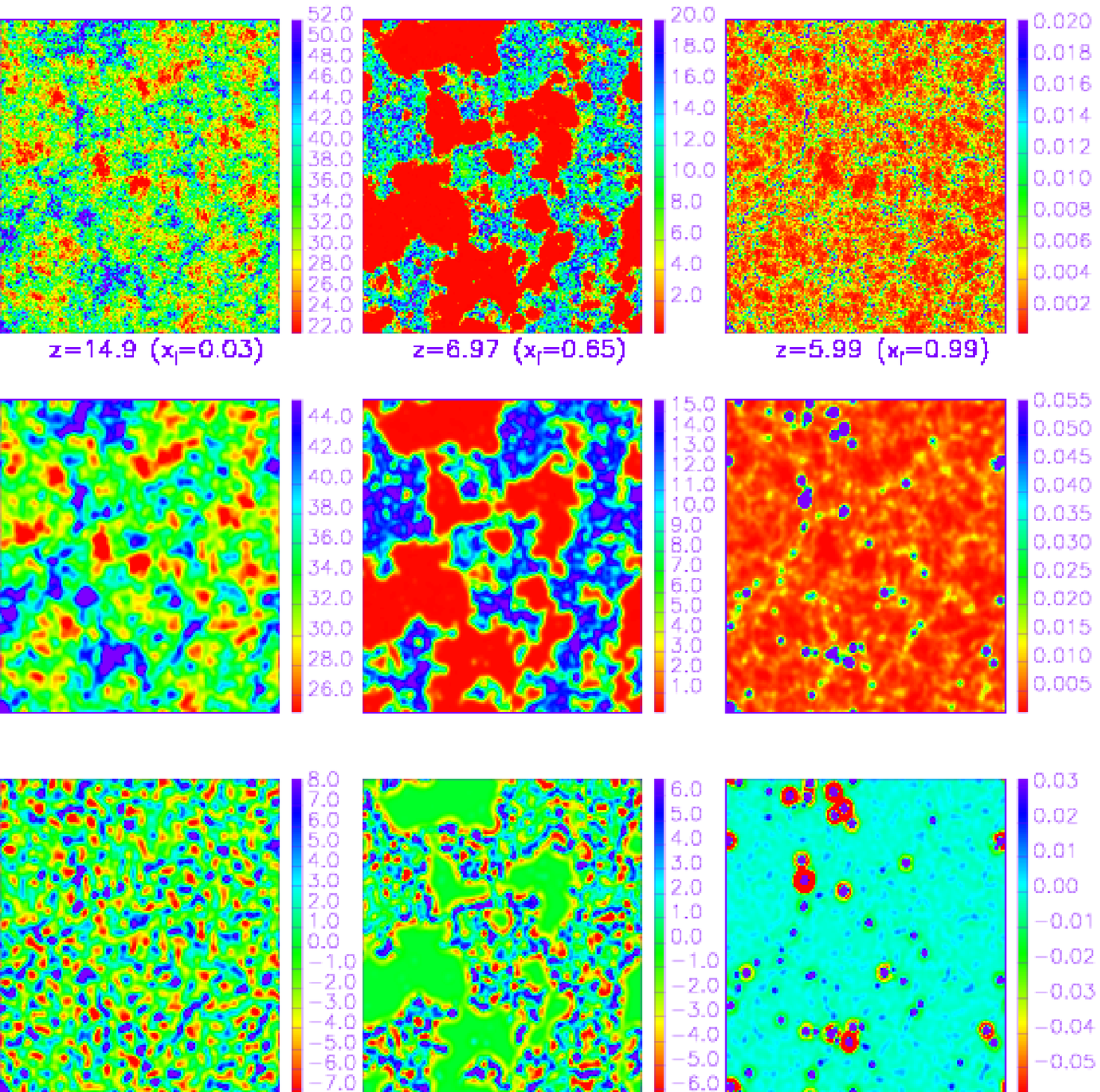}}
\subfigure[HRT sim 2]{\includegraphics[width=0.45\textwidth]{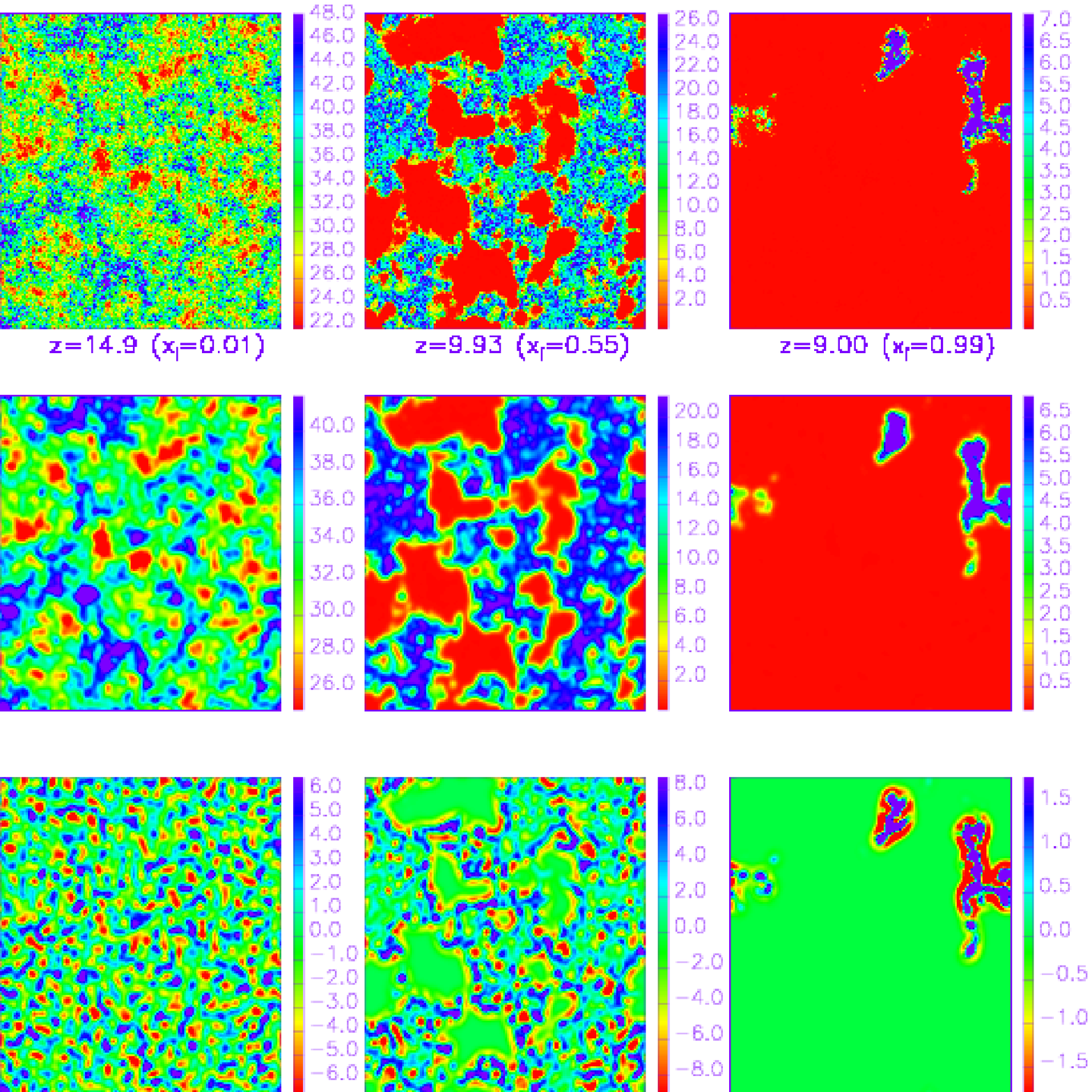}}\\
\subfigure[21cmFAST1]{\includegraphics[width=0.45\textwidth]{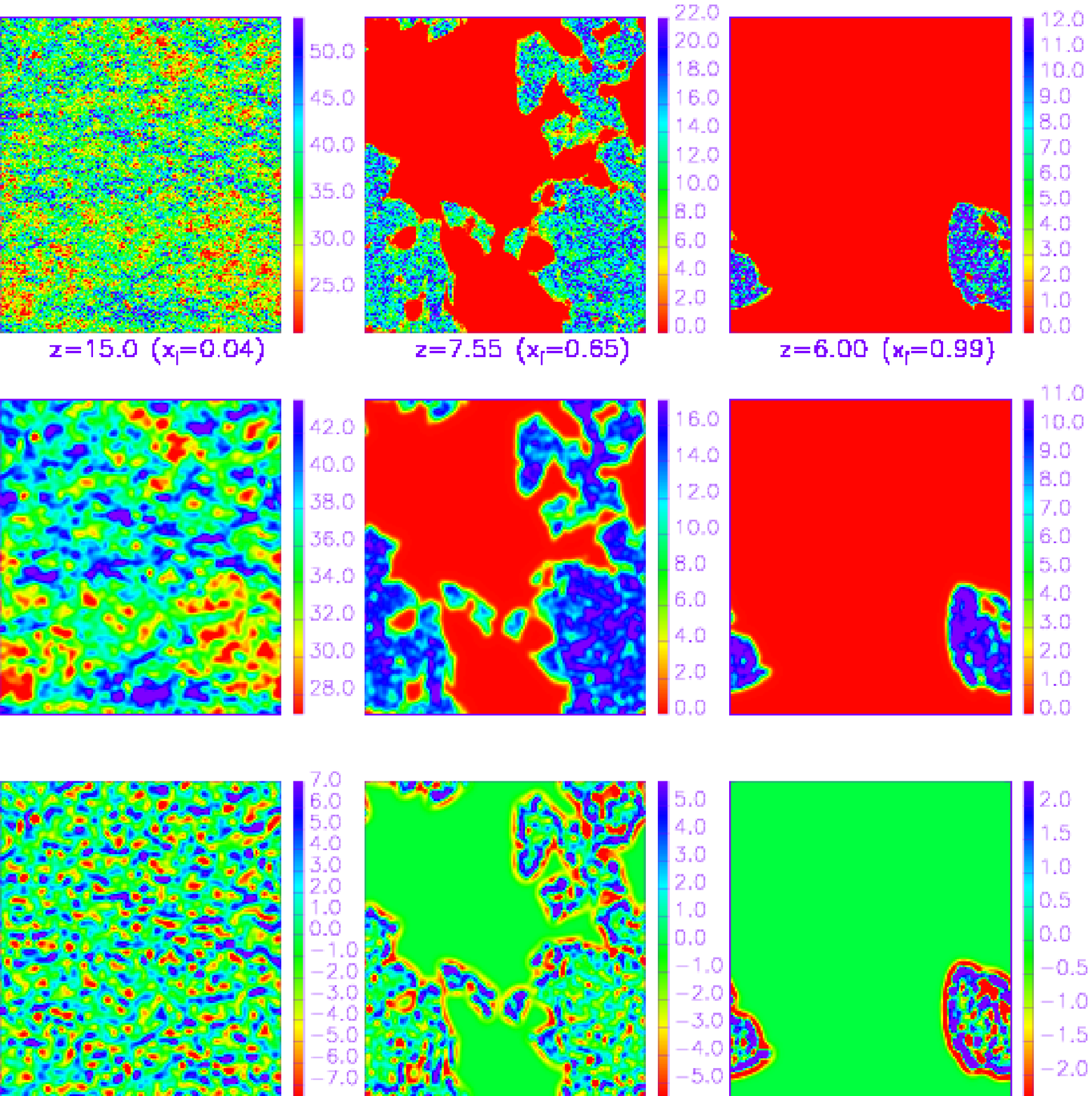}}
\subfigure[21cmFAST2]{\includegraphics[width=0.45\textwidth]{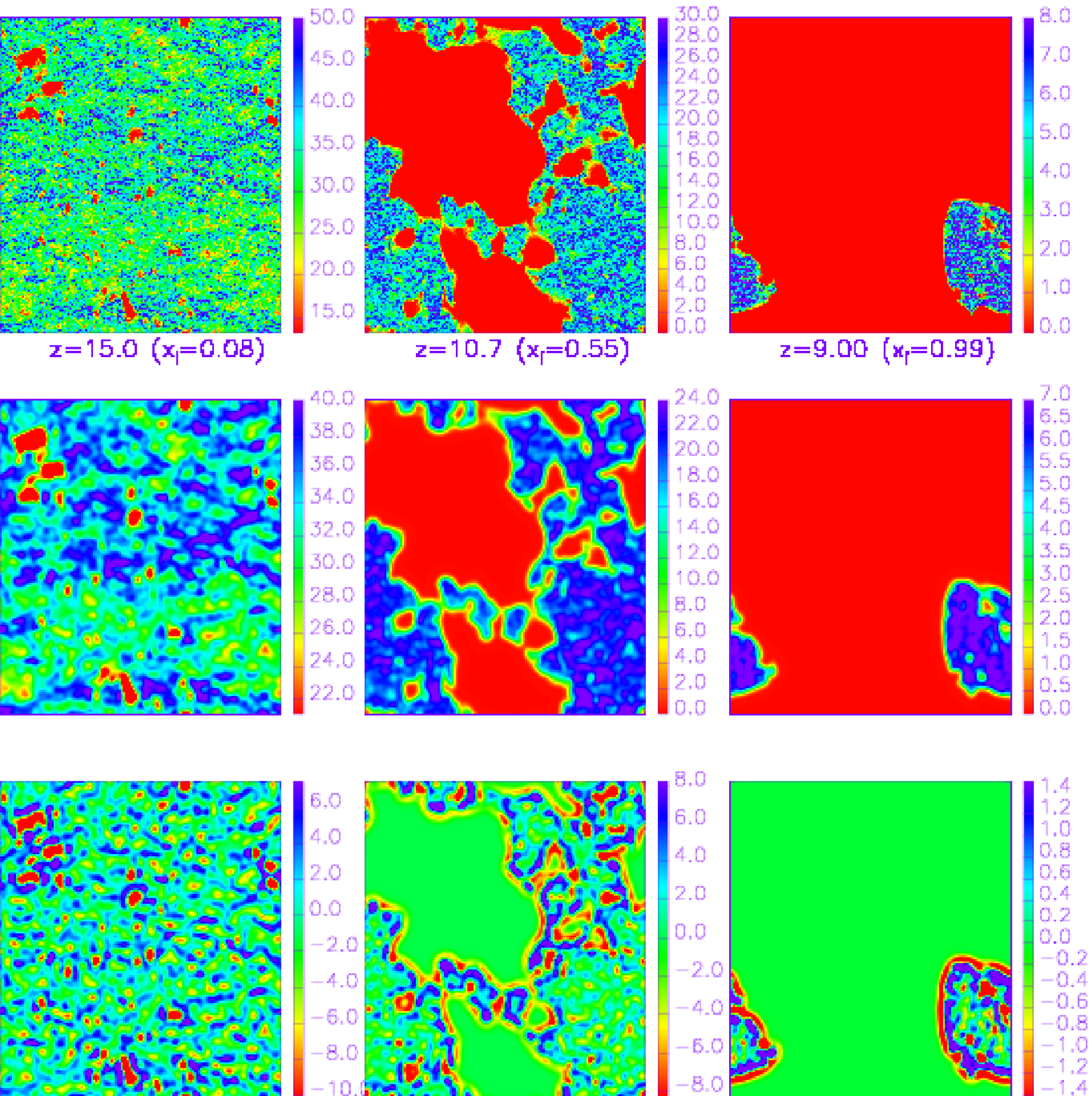}}\\
\caption{The 21cm maps for the four models: (a) HRT sim1, (b) HRT sim2, 
(c) 21cmFAST1, (d) 21cmFAST2. For each model, three redshifts are plotted (marked on
the figure). Also, for each model, the 
Top panel shows the  original $\delta T_b$ map with frequency bin width 
$\Delta\nu$ = 0.2 MHz. The high and low differential brightness temperature regions are the neutral and ionized ones
respectively.The unit of the color bar is mK. The Middle and bottom panels show the simulated ``observed map'' 
with the Gaussian and compensated Gaussian beams respectively, 
with beam width $\Delta\theta=1^{'}$.}
\label{fig:map}
\end{figure*}

 In Fig.\ref{fig:map} we show the evolution 
of 21cm maps at the early, middle and late stage of reionization
for the four models: (a) HRT sim1, (b) HRT sim2, 
(c) 21cmFAST1, (d) 21cmFAST2.  
In each of the subfigures,  the top panels represent the 
original $\delta T_b$ signal, the middle and bottom panels
are the simulated observed map with the Gaussian and compensated 
Gaussian beam profile, respectively.   
We can see that there even though both the HRT sim1 and 21cmFAST1 models
have relatively late reionization, there are significant differences 
between the two models at each epoch. In the
HRT sim1, the ionized regions are diffused, while it is linked 
together in 21cmFAST1. Similarly, for the two 
early reionization models (HRT sim2 and 21cmFAST2), there are also 
distinctive differences. This 
indicates that the reionization process 
in the HRT simulation and that in the 21cmFAST are different. For the two 
21cmFAST simulations, the distributions of $\delta T_b$ are similar if 
they have the same ionization fraction $x_{i}$. Since the Gaussian and compensated Gaussian beam
can smooth the $\delta T_b$ distribution, the largest value of the $\delta T_b$ decreases after the smoothing. 
As explained above,  the compensated Gaussian beam can produce negatives values, therefore, $\delta T_b$ 
is negative in some regions with the compensated Gaussian beam.

\begin{figure}
\resizebox{\hsize}{!}{\includegraphics{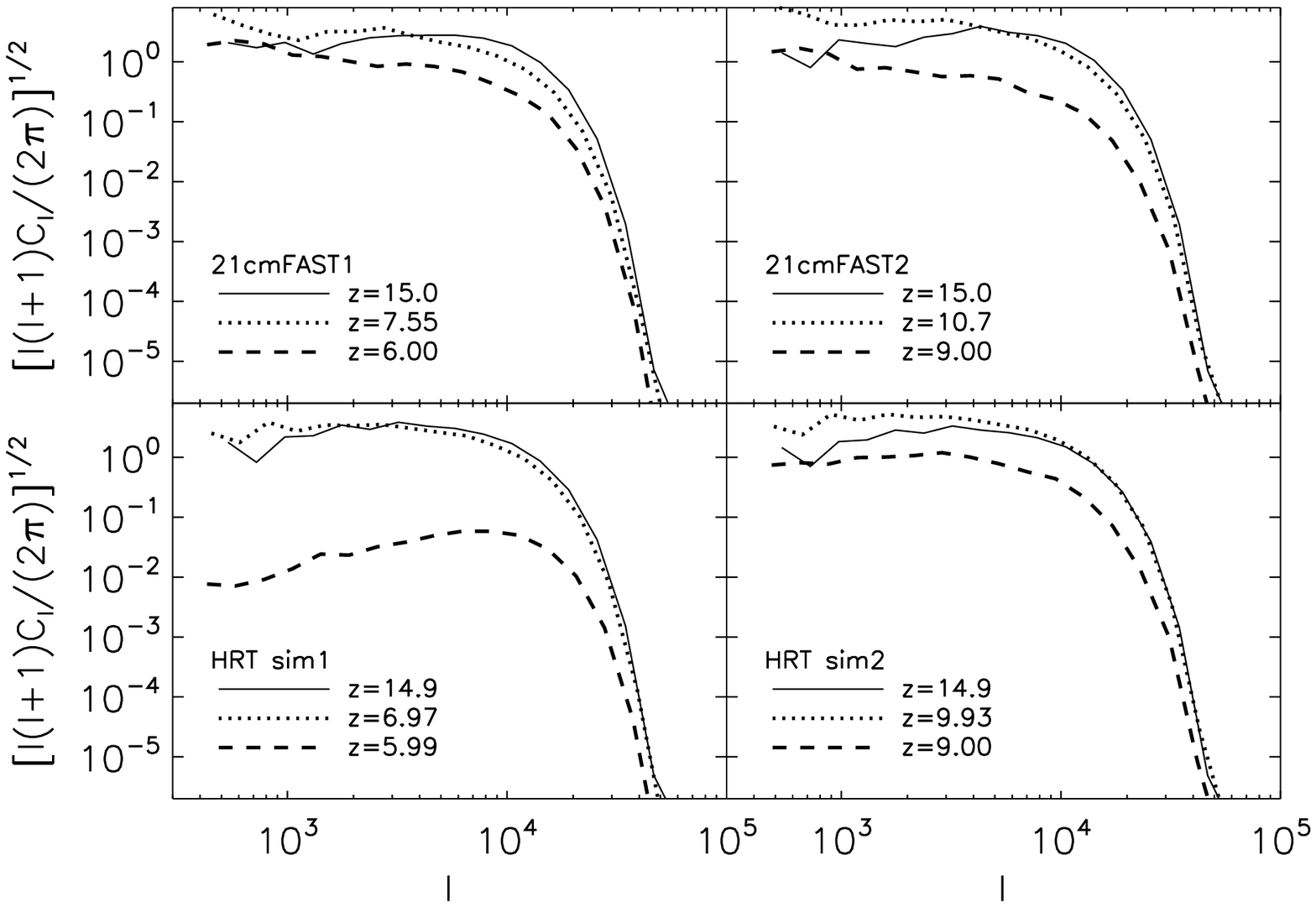}}
\caption{2d angular power spectrum of $\delta T_b $ for four models with different redshifts.}
\label{fig:pow}
\end{figure}

In order to compare the topology results with  those from 
the  power spectrum,   we also calculate the angular power spectrum
$[l(l+1)C_l/2\pi]^{1/2}$. 
In Figure~\ref{fig:pow} we plot the angular power spectrum of $\delta Tb$ 
for the four models with different redshifts.  
It is seen that the shape of the angular power spectrum is nearly the same 
for the early and middle phase of reionization in the four simulations.  
Except for HRT sim1, the shape of the angular power spectra in the late 
phase of reionization are also similar, hence it would be difficult to 
distinguish the different reionization scenarios from the angular
power spectrum, while the topology offers a way to distinguish them.

\begin{figure*}[p]
\subfigure[HRT sim 1]{\includegraphics[width=0.45\textwidth]{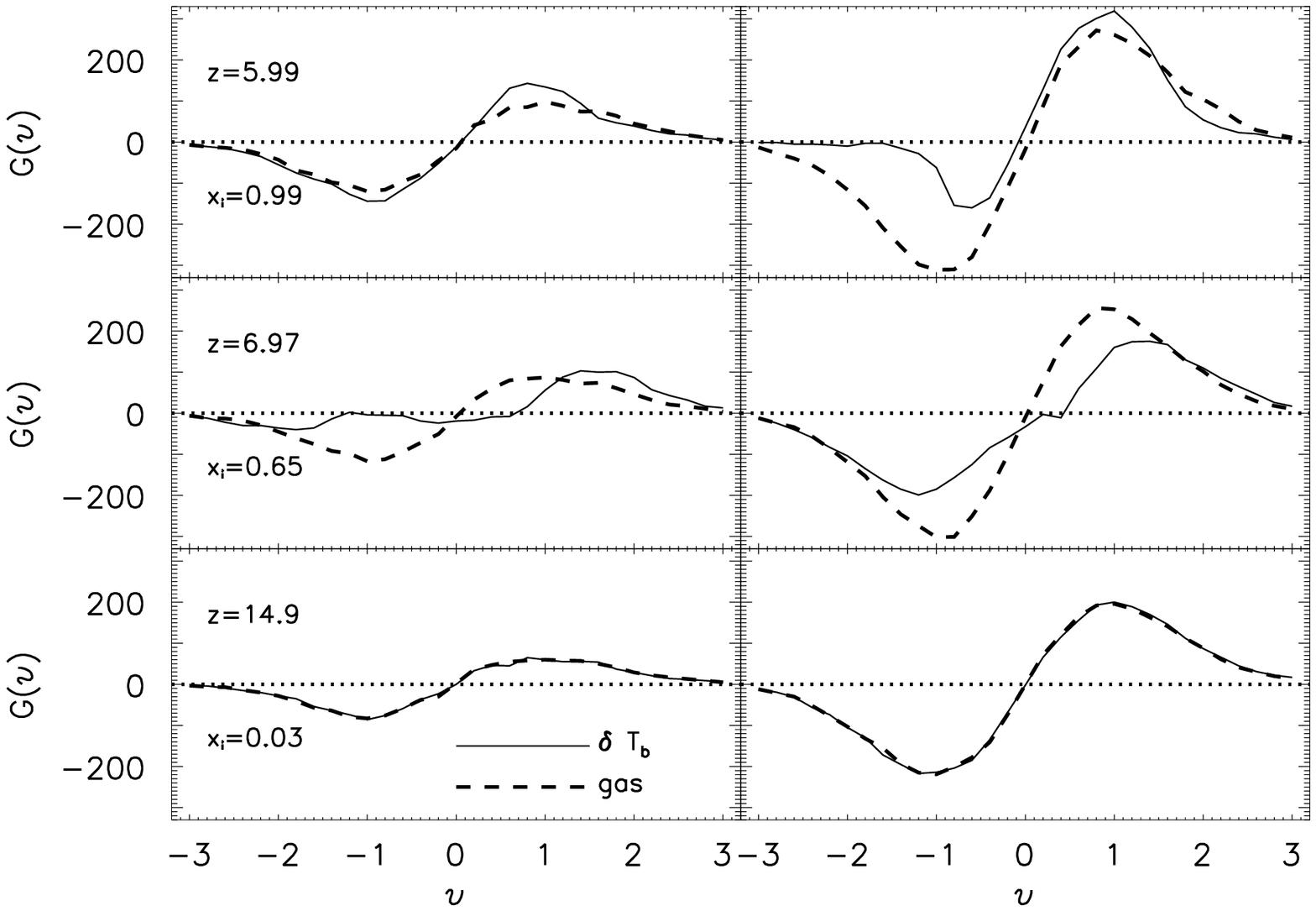}}
\subfigure[HRT sim 2]{\includegraphics[width=0.45\textwidth]{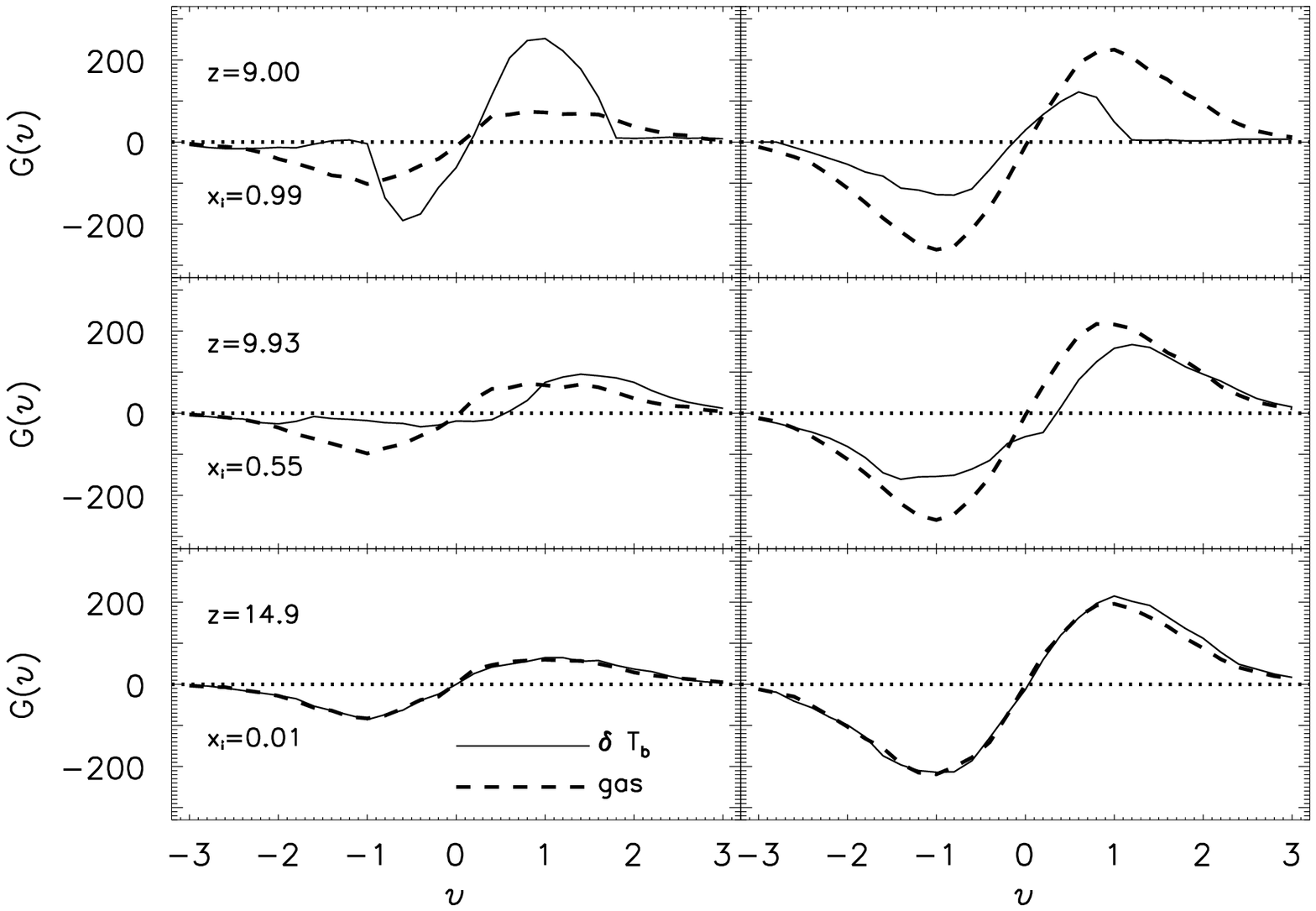}}\\
\subfigure[21cmFAST1]{\includegraphics[width=0.45\textwidth]{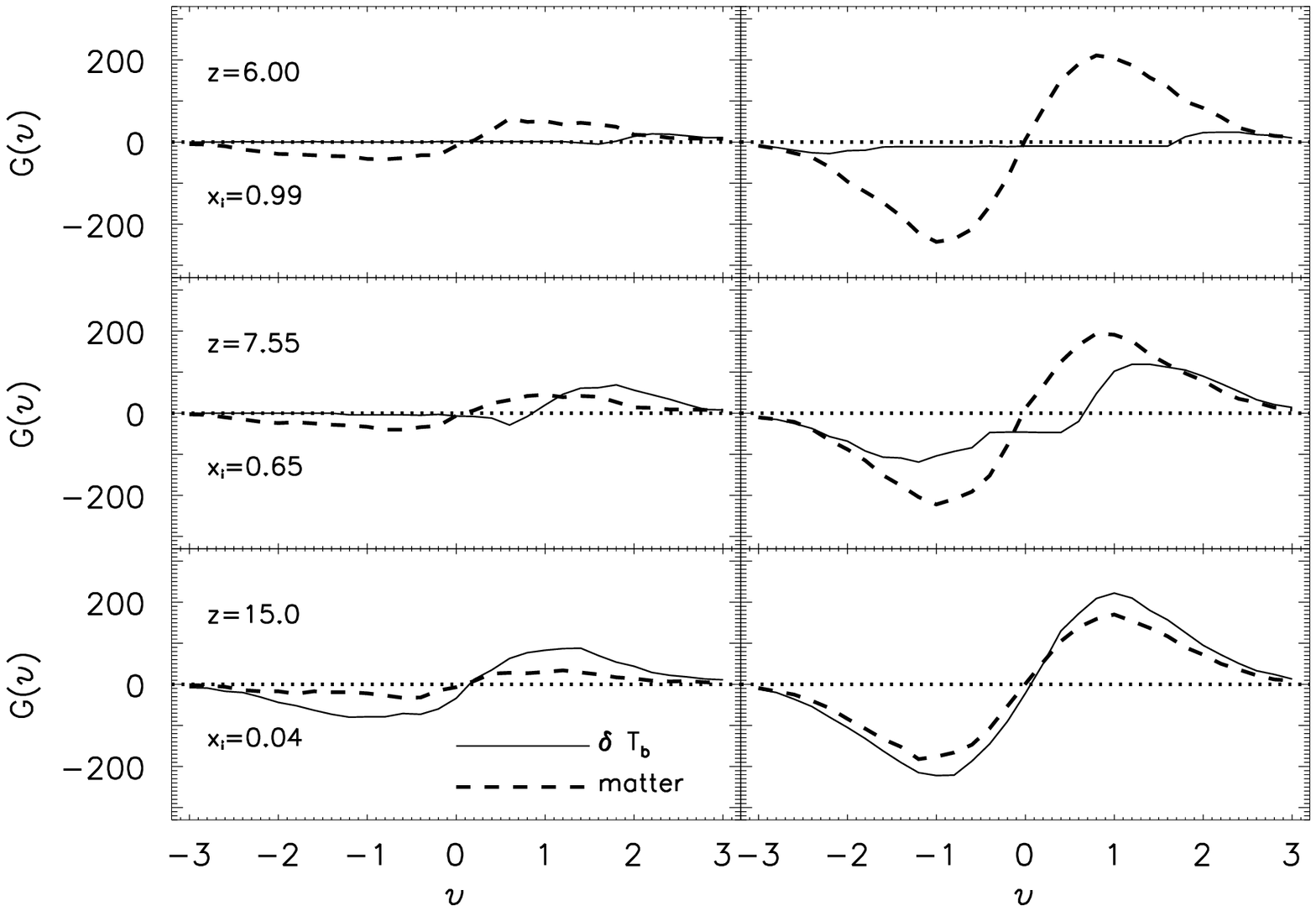}}
\subfigure[21cmFAST2]{\includegraphics[width=0.45\textwidth]{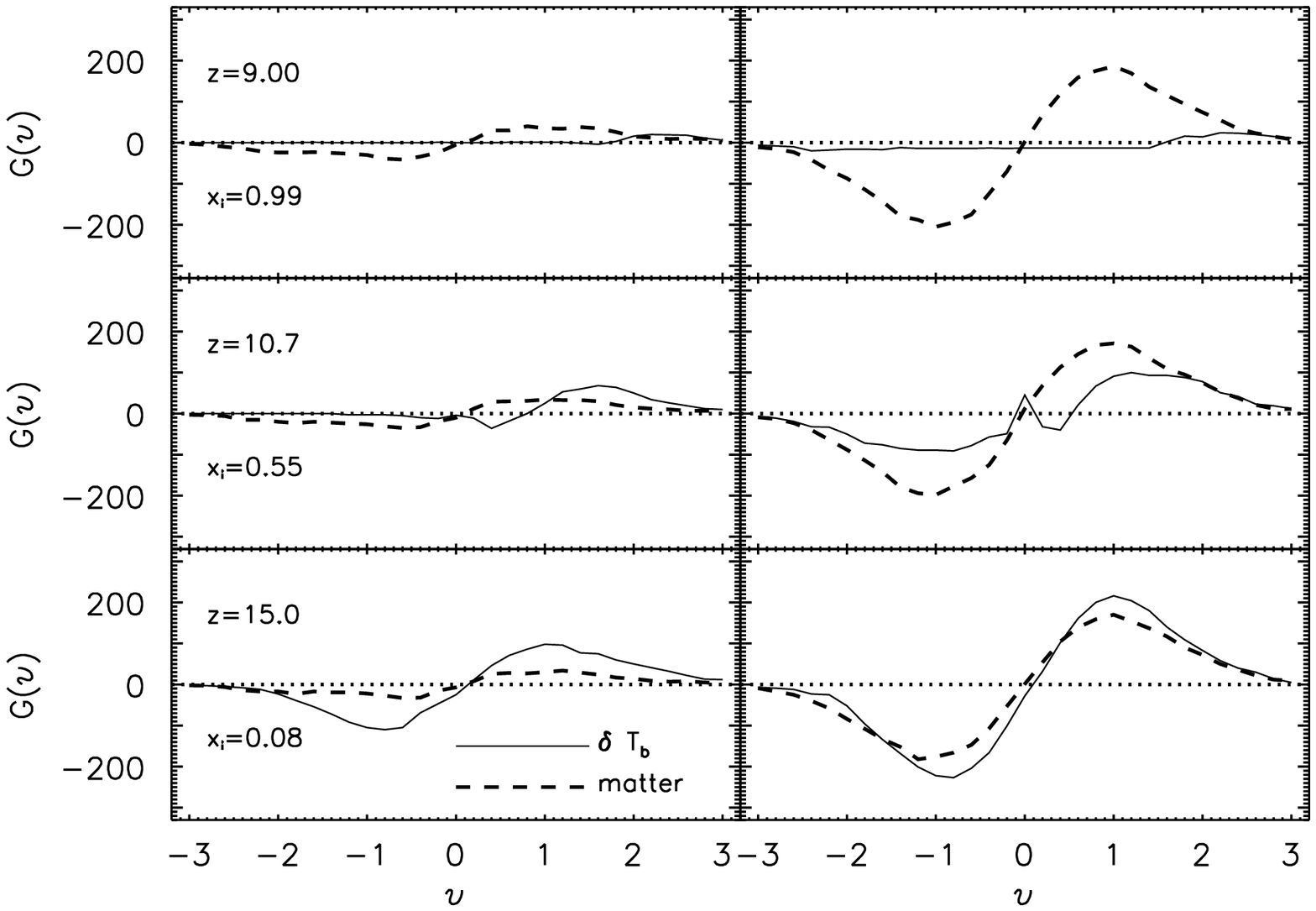}}\\
\caption{The 2d genus of $\delta T_b$ (solid) and gas distribution (dashed) 
at different redshifts for  (a) HRT sim1, (b) HRT sim2, 
(c) 21cmFAST1, (d) 21cmFAST2 models. In each model, the left panels show the results for 
Gaussian beam while the right ones represent results for compensated 
Gaussian filter.  
}
\label{fig:2dgenus}
\end{figure*}

In Figure~\ref{fig:2dgenus}, we 
show the 2d genus for both the differential brightness temperature (solid lines) 
and the gas (or matter) density (dashed lines) in the four simulations. On the 
scale discussed here, the gas (or matter) density distribution is nearly 
Gaussian, so from the comparison between the density and  $\delta T_b$ 
genus curve, we can see how far do the 21cm signal 
deviates from Gaussian. In each figure, the left and right 
panels show the results from the Gaussian beam and compensated 
Gaussian beam, respectively. Since Panel (a) and (b) belong to the same kind of simulation, 
and Panel (c) and (d) belong to the same kind of simulation,  and the ionization fraction are also similar, 
the genus curve on the left and right panels look similar.

It is noted that the 2d genus curve from the Gaussian beam 
is virtually indistinguishable from the compensated Gaussian one.
The  compensated Gaussian beam has a Gaussian-type peak in the middle, surrounded by
a negative wing, which adds small wiggles to the real signals. Therefore, the compensate Gaussian beam change the
topology of the differential brightness temperature slightly  ~\citep{2014JKAS...47...49H}. More detailed 
comparison of these two beams are shown in  \cite{2014JKAS...47...49H}.
In this paper we only present the results for the two beam profiles here, 
and from now on we will only discuss the results obtained with the Gaussian beam. 

For each model, the 2d genus curve of the differential brightness temperature is
distinctly different at different ionization fractions or redshifts. Furthermore, 
even at the same ionization fraction, 
the 2d genus curves for different reionization models are significantly different.
This can be seen clearly from the middle panels in Fig.~\ref{fig:2dgenus}. Quantitatively, we can make a
Kolmogorov-Smirnov (KS) test, which is used to test whether two distributions are different. Usually, the KS-test gives the deviation between two probability distribution functions of a single independent variable, 
but it is also valid to distinguish two any arbitrary distributions, that is, the multivariate distribution. Here we take $G(\nu)$ as the distribution function of the threshold $\nu$, 
and apply the KS-test to the  $G(\nu)$ functions. Note that we are not testing the distribution of the 21 cm brightness temperature, but comparing the shape of the genus curves $G(\nu)$, taking it 
as if it the distribution of the single variable $\nu$. The significance level of an observed value of $D$, which is a disproof of the null hypothesis that the distributions are the same, is given by  ~\citep{1992nrfa.book.....P} :
\begin{equation}
prob(D>observed)=Q_{KS}\bigg(\bigg[\sqrt{N_e}+0.12+0.11/\sqrt{N_e}\bigg]D \bigg)
\end{equation}    
where $D$ is defined as the maximum value of the absolute difference between two cumulative distribution functions, $N_e=\frac{N_1N_2}{N_1+N_2}$ ($N_1$ and $N_2$ are the data number in distribution 1 and 2, respectively), and the
function $Q_{KS}$ is defined as
\begin{equation}
Q_{KS}(\lambda)=2\sum_{j=1}^{\infty}(-1)^{j-1}\exp(-2j^2\lambda^2)
\end{equation}
 Small values of $prob$ imply that the distribution 1 is significantly different from that of distribution 2. 
We refer the interested reader to the book ~\citep{1992nrfa.book.....P} for the detail description of the KS test.              
The result shows that  the genus curve from HRT sim1 at $x_i=0.65$ is different from
that from 21cmFAST1 at $x_i=0.65$ with confidence level higher than $97\%$.

In the early phase of reionization, i.e. $x_{i}\le0.08$, the universe is 
nearly neutral, the differential brightness temperature of HI, $\delta T_b$,  
still follows the Gaussian distribution, which can be seen from the bottom 
left panels in the subfigures of Fig.~\ref{fig:2dgenus}.  
The amplitude of the $\delta T_b$ genus curve is larger than 
the gas  density genus curve at $\nu\sim -1$ 
at $z=14.9 $ in HRT sim1.  The major reason for this is 
that the star formation is already occurring during this epoch, and some 
regions have already been ionized, which increase the number of low-density
regions.  
In simulations 21cmFAST1 and 21cmFAST2, the 2d genus curve of $\delta T_b$ 
is distinguishable from the matter curve:  the amplitude of the 2d genus curve of $\delta T_b$ 
is larger than that of the matter from low $\nu$ to high $\nu$.  This is the result of 
a combination of two reasons. First, similar to the HRT sim1 and HRT 
sim2 models, the universe has begun to ionize at this redshift, 
both new islands of HI regions and the lake 
of HII regions  have formed, the former corresponds to the isolated high-density regions, 
while the later is responsible for the low-density regions, the same findings were also 
presented in  ~\cite{2014JKAS...47...49H}. 
Second, the value of  $\delta T_b$ is 
related to the comoving velocity gradient of gas along the LOS ${\rm d}v_r/{\rm d}r$,  
see Eq.~(\ref{eq_tb2}). We  find that ${\rm d}v_r/{\rm d}r$ in simulation HRT sim1 
and HRT sim2 is tiny, while those in simulation 21cmFAST1 and 21cmFAST2 are 
relatively large.    
 
In the middle phase of reionization, i.e., $0.55\le x_{i}\le 0.65$ , many bubbles 
exist and some of them overlap, therefore, the amplitude of genus curve of $\delta T_b$ 
is nearly zero at low $\nu$ and the genus curve of $\delta T_b$ is shifted to the 
right compared with the genus curve of 
the gas (or matter). This shift is  consistent with the result in Figure 2 
of ~\cite{2007ApJ...663..244K} . In their studies,  the genus curve for a uniform 
disk with randomly distributed empty holes shifts to the right.  The amplitude of the 
2d genus curve of $\delta Tb$ in the high $\nu$ still keeps the vestiges  of its 
initial curve. This is because the high-density regions are still shielded from 
the ionized photons to ionize the rest of the universe ~\citep{2008ApJ...675....8L}.
 
In the late epoch of reionization, i.e., $x_{i}\leq0.99$, the universe is 
almost completely ionized, the genus of $\delta T_b$ from HRT sim1 shows that the remaining
HI nearly follows the matter distribution, which agrees with the result given
in \cite{2008ApJ...675....8L}.   Although $x_{i}=0.99$ at redshift $z=8.9954$
 in HRT sim2, there are still some neutral patches, 
and the amplitude of the genus curve for $\delta T_b$ is higher than that of the
 matter distribution.  For the two 21cmFAST simulations, the ionized  bubble has
merged except for very low density regions, therefore, the genus 
curves of $\delta T_b$ at low $\nu$ is nearly zero.

\begin{figure*}[p]
\subfigure[HRT sim 1]{\includegraphics[width=0.45\textwidth]{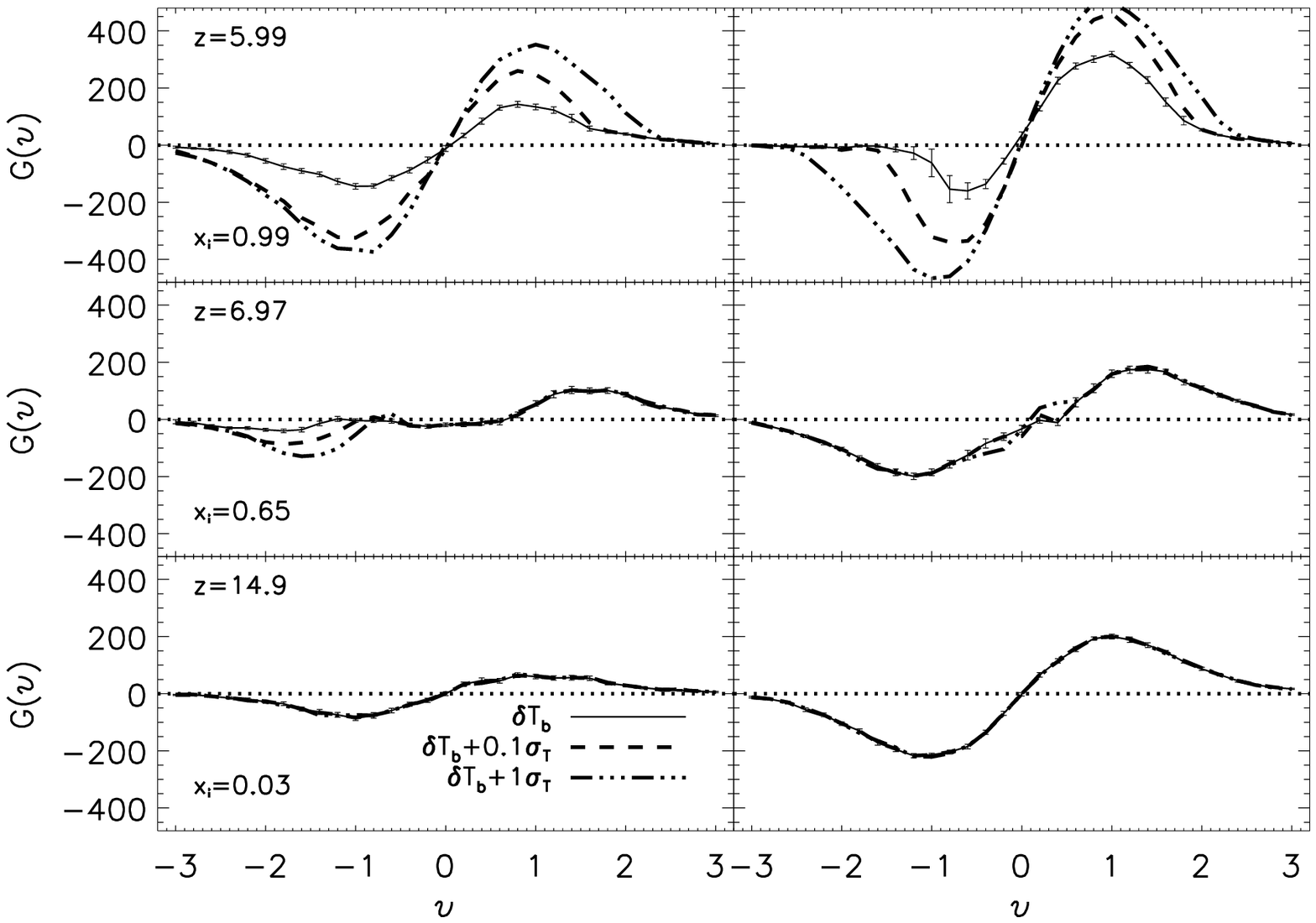}}
\subfigure[HRT sim 2]{\includegraphics[width=0.45\textwidth]{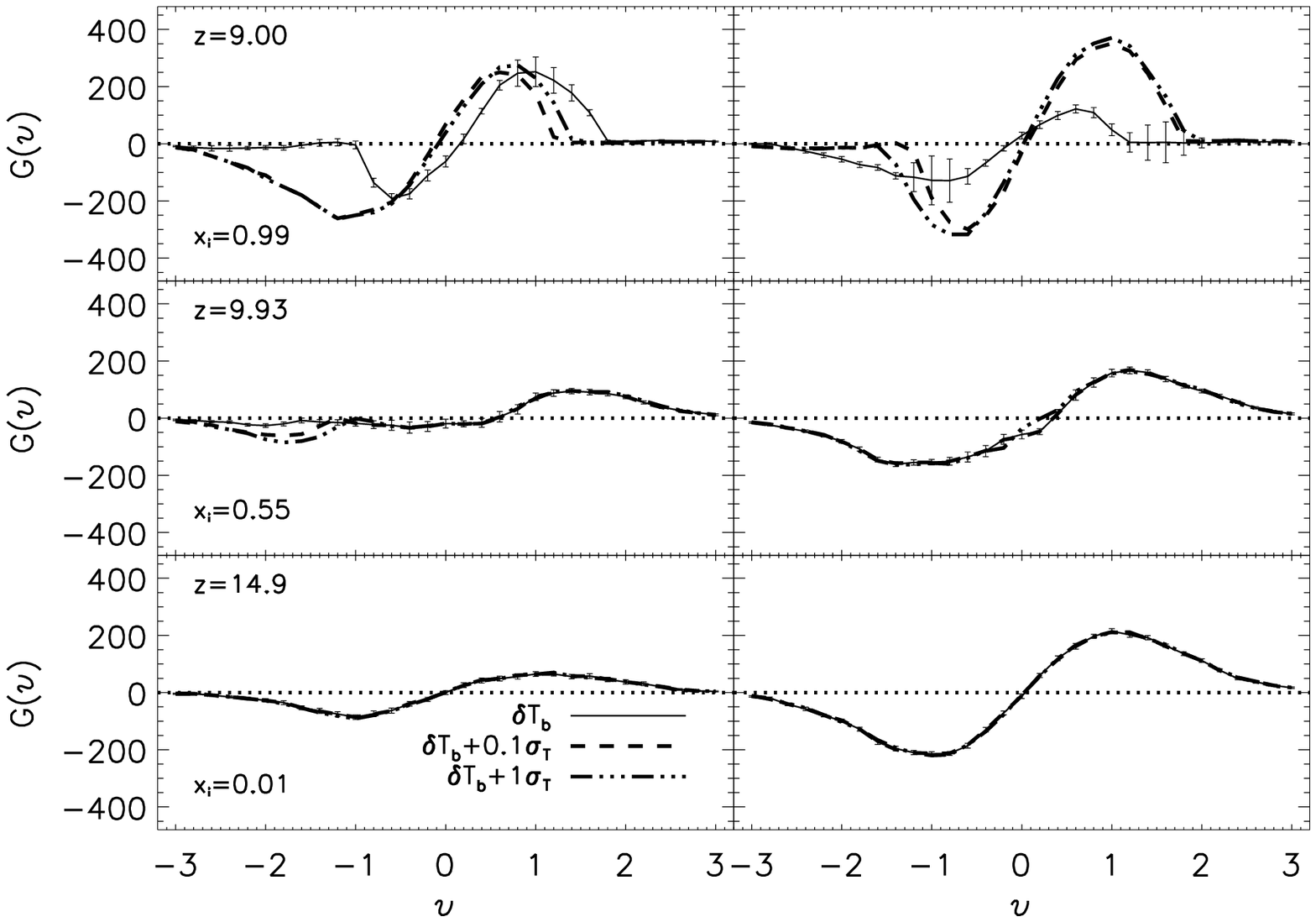}}\\
\subfigure[21cmFAST1]{\includegraphics[width=0.45\textwidth]{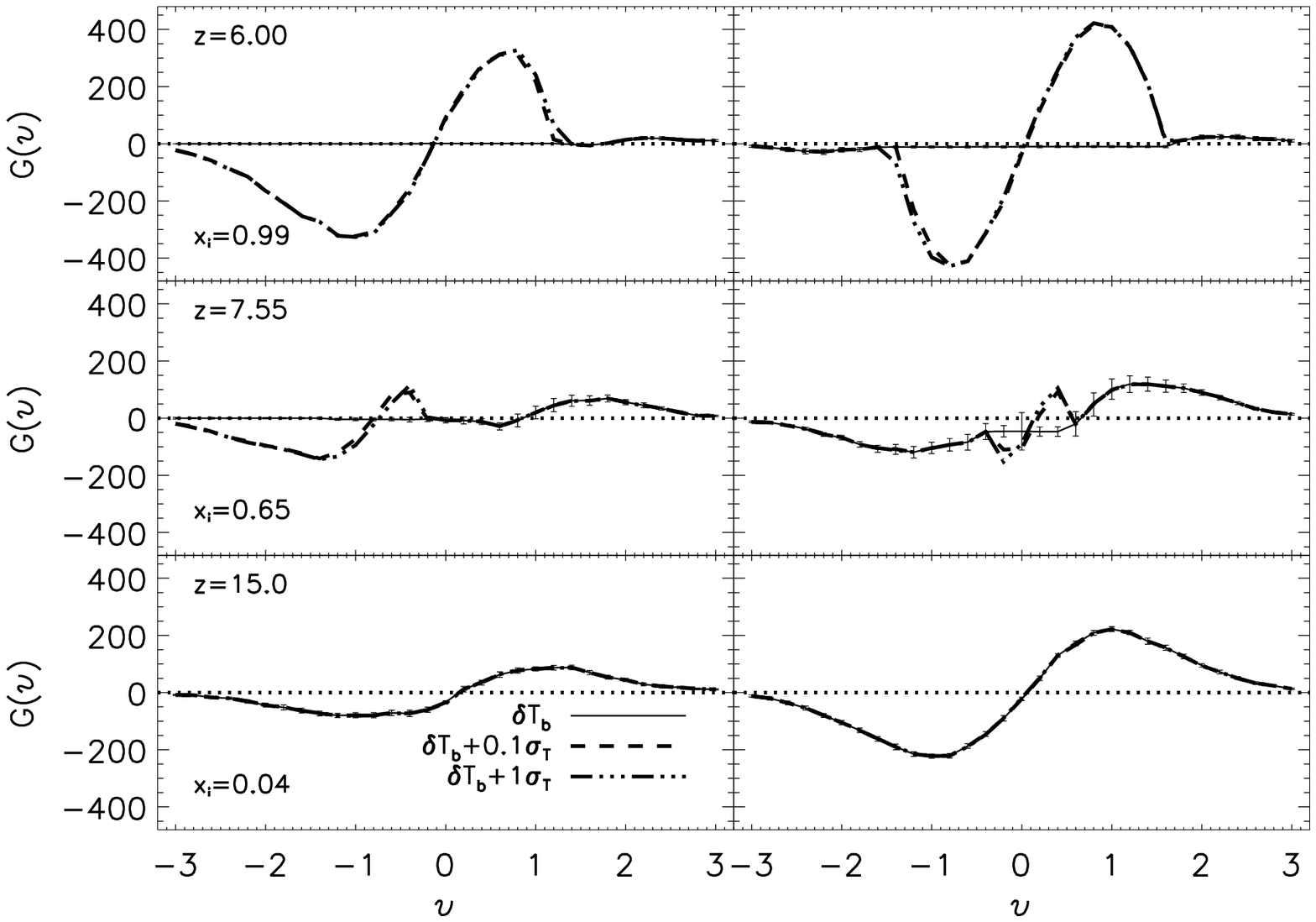}}
\subfigure[21cmFAST2]{\includegraphics[width=0.45\textwidth]{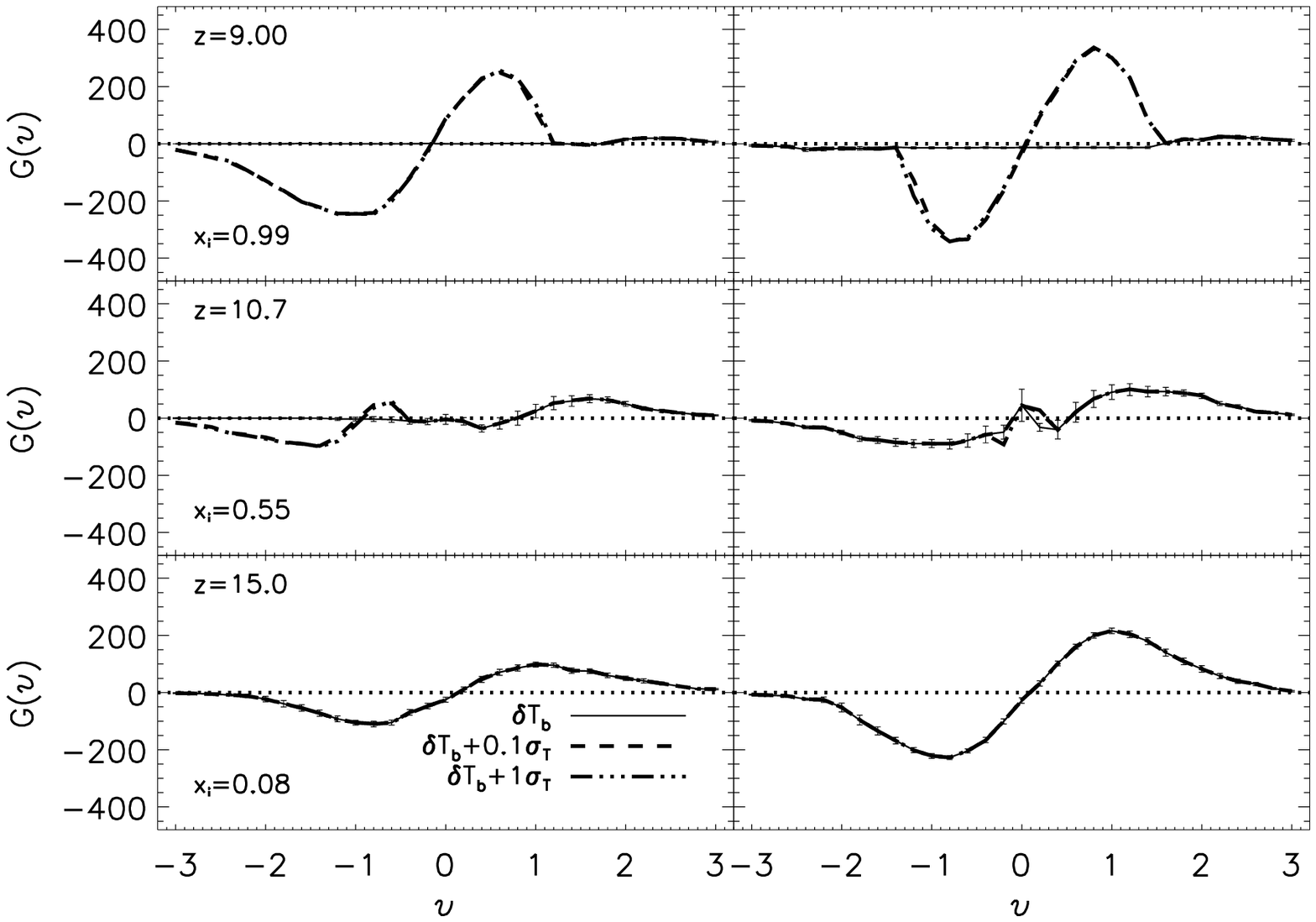}}\\
\caption{Similar to Figure~\ref{fig:2dgenus}, but with Gaussian noise added. In each panel, 
the solid line represent the genus curve
without the thermal noise, while the dashed and dash-dot lines represents the results with $0.1\sigma_T$ and $\sigma_T$ thermal noise added, respectively. 
The error bar in the solid line is estimated by 30 similar simulated data samples.} 
\label{fig:2dgenus_noise}
\end{figure*}

In Figure~\ref{fig:2dgenus_noise}, 
we show the effect of the the thermal noise on the genus curve. It is seen that the genus curve is not affected in the early phase of reionization. 
The reason is that   the brightness temperature  $\delta T_b$ follows the Gaussian distribution, when an additional Gaussian  noise is included, the distribution of $\delta T_b$
is still Gaussian.  In the middle and late epoch of reionization, the effect of the thermal noise on the genus curve is significant. 
The amplitude of the genus curves with thermal noise are larger than those without thermal noise due to merging of bubble.
This effect is in some sense similar to what was shown in Fig.5 of \cite{1989ApJ...345..618M}, where the amplitude of the genus curve is decreased after structure formation. 
Nevertheless, as demonstrated by the KS test, the two ionized models can still be distinguished clearly even when $1\sigma_T$ noised is added. 

\begin{figure}
\resizebox{\hsize}{!}{\includegraphics{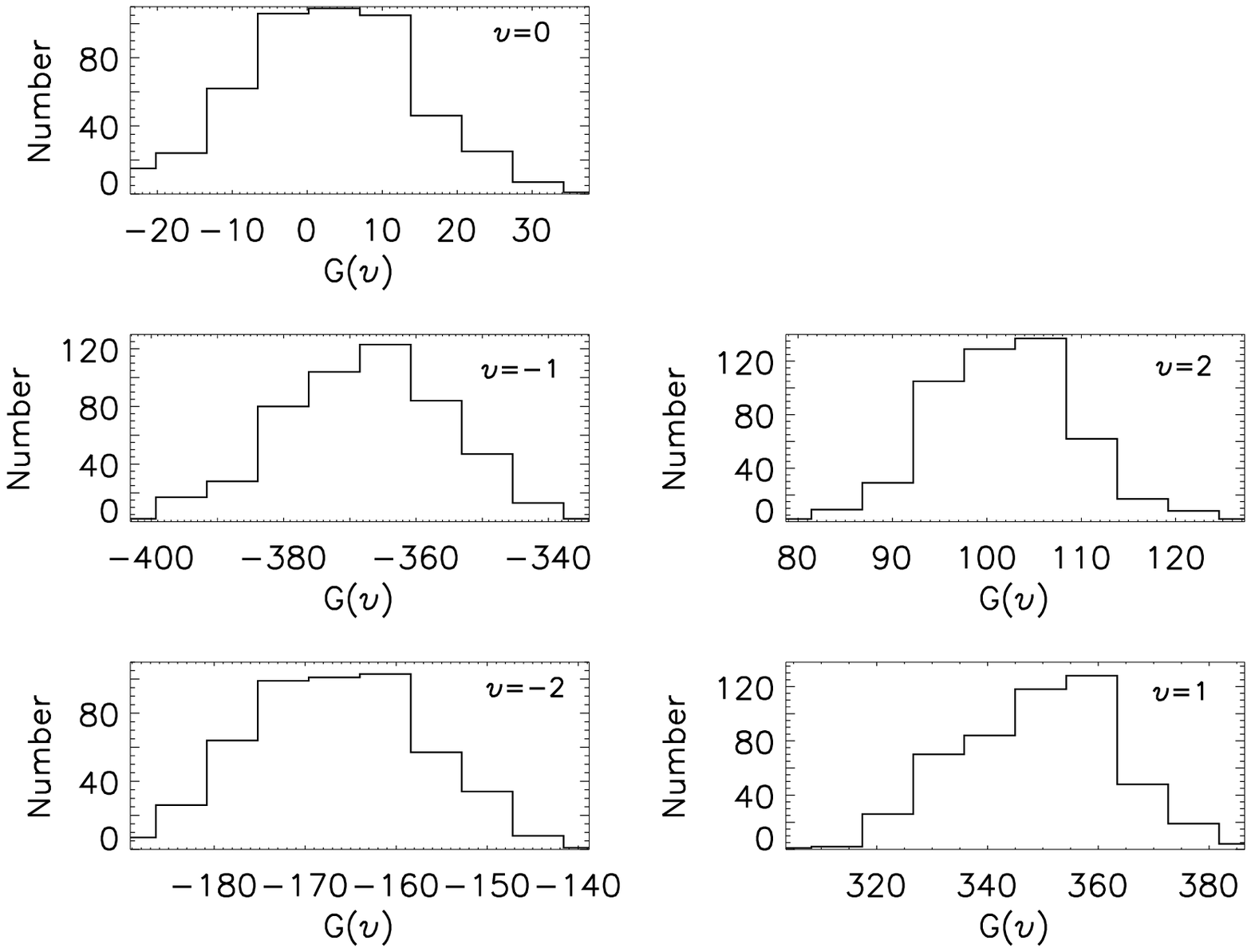}}
\caption{Probability density function of the 2d genus including $1\sigma_T$ Gaussian noise  for HRT sim1 at $z=5.99$ for $\nu$=-2, -1, 0, 1, 2.}
\label{fig:pdf_genus}
\end{figure}

In Figure~\ref{fig:pdf_genus}, the PDF of the 2d genus $G(\nu)$ at $\nu$=-2, -1, 0, 1, 2.  are plotted for 
the   the HRT sim1 model at $z=5.99$ with $1\sigma$ thermal noise added.  
These PDFs of the 2d genus show that they are centered at certain value, with nearly Gaussian distribution. Thus, in making model test it is reasonable to assume Gaussian likelihood for the genus measurement.

\begin{figure*}[p]
\subfigure[angular power  spectrum]{\includegraphics[width=0.45\textwidth]{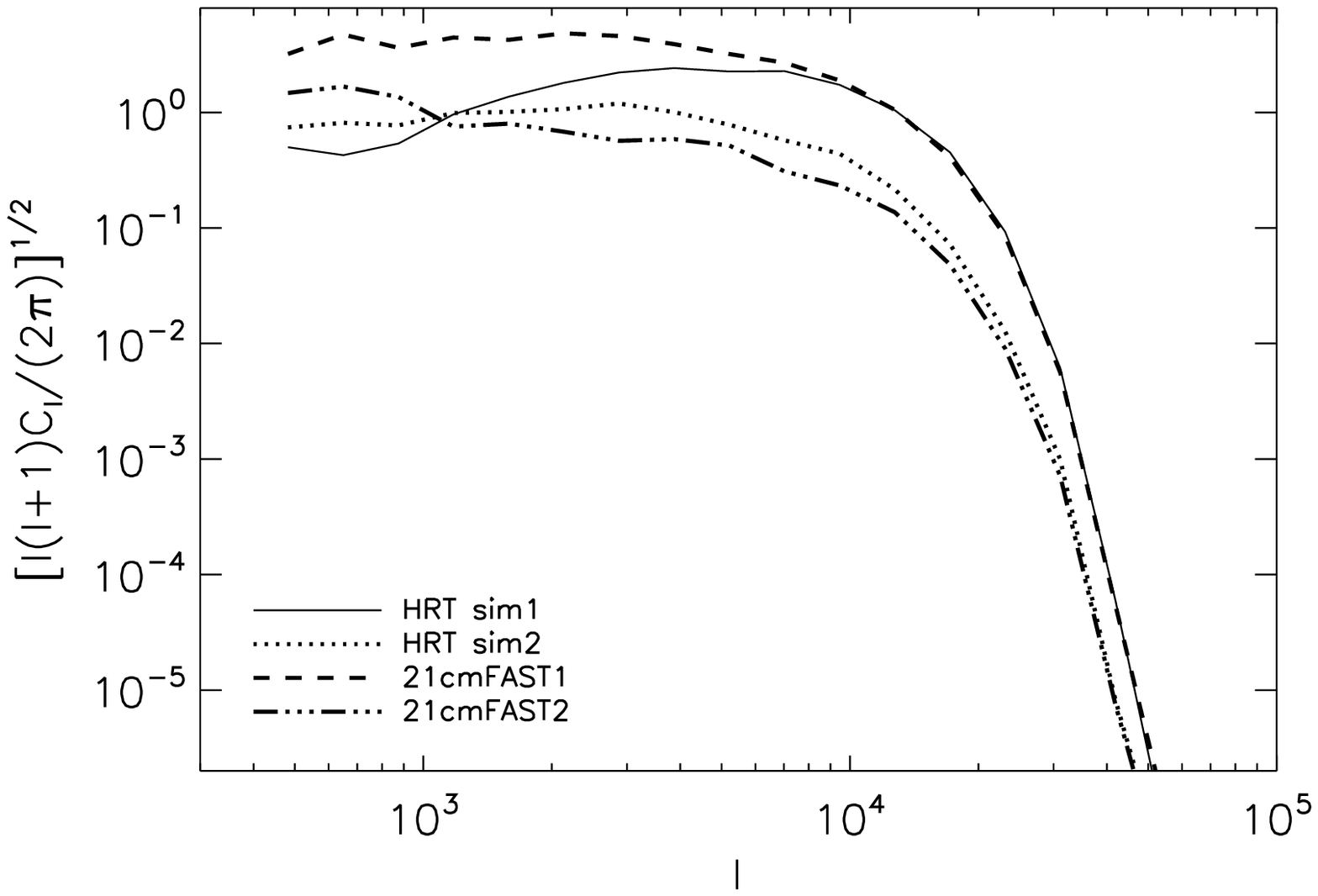}}
\subfigure[2d genus curve]{\includegraphics[width=0.45\textwidth]{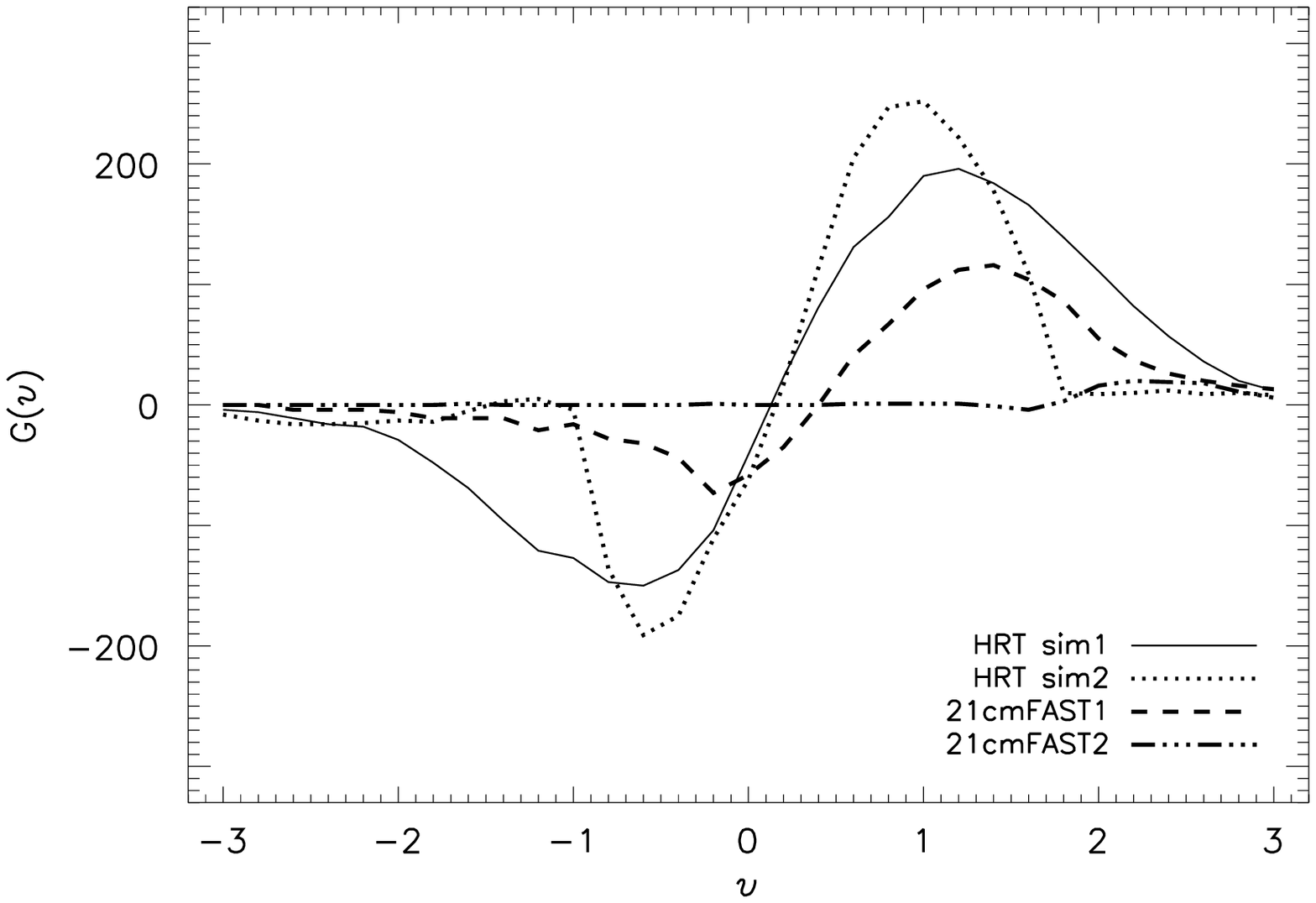}}\\
\subfigure[angular power spectrum with $1\sigma$ thermal noise ]{\includegraphics[width=0.45\textwidth]{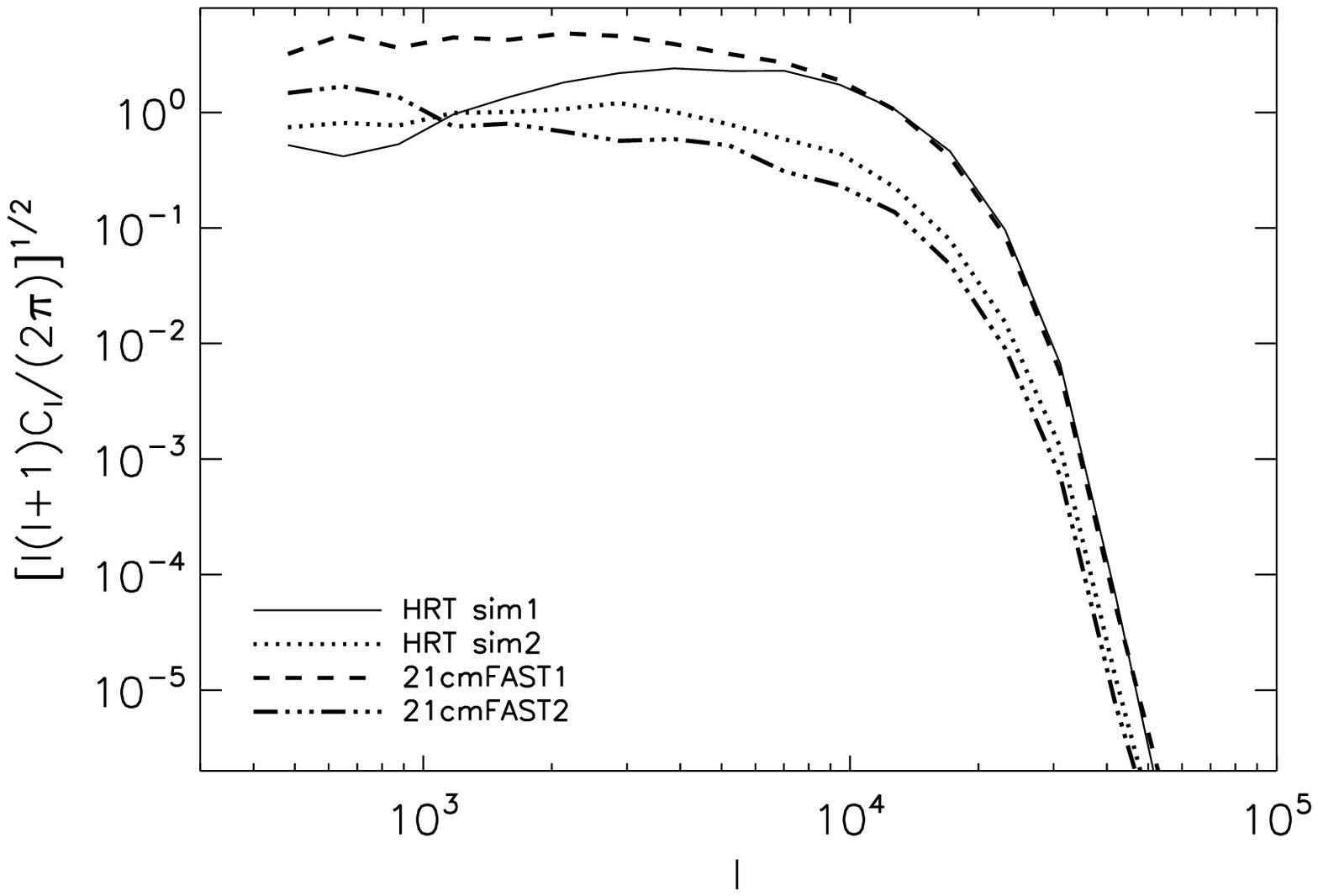}}
\subfigure[2d genus curve with $1\sigma$ thermal noise]{\includegraphics[width=0.45\textwidth]{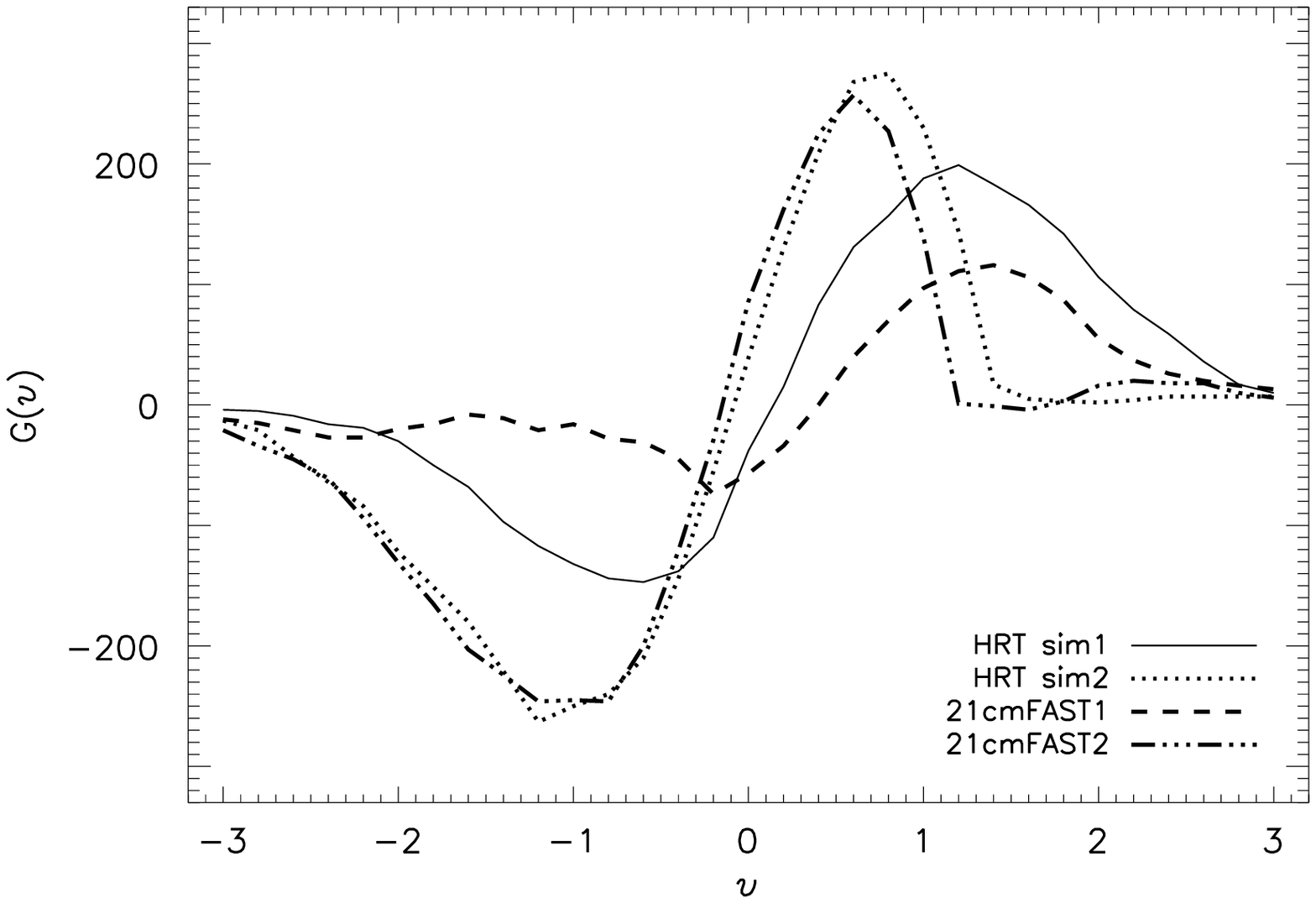}}\\

\caption{(a) 2d angular power spectrum of $\delta T_b$ for four models with the same redshift $z=9.00$.
(b) 2d genus distribution of $\delta T_b$ for four models with the Gaussian beam with the same redshift $z=9.00$. (c) similar to (a), but for the results with 1$\sigma$ thermal noise. (d) similar to (c), but for the results with 1$\sigma$ thermal noise.}
\label{fig:pow_genus}
\end{figure*}

In Figure~\ref{fig:pow_genus}, we compare the 2d angular power spectrum (left panels) and the 2d genus curve (right panels) for the four different reionization models at the same redshift $z=9.00$ with and without $1\sigma$ thermal noise.  We can also obtain some statistical 
confidence level from these tests--the different models can be compared using the KS test with the angular power
spectrum data and the 2d genus curves (See Table~\ref{tab:prob}) .  Here we take the angular power spectrum as the distribution function of the single variable $l$. From the value of $prob$, we know that the 21cmFAST1 model can be better distinguished with the other three 
models by using the 2d angular power spectrum, while the 21cmFAST2 model can be better distinguished by using the 2d genus curve if there is no thermal noise. 

The Gaussian thermal noise does not affect the angular power spectrum,  but it can reduce the non-Gaussian signal of the 2d genus. From the lower part of Table~\ref{tab:prob},  we see that the HRT sim2 model can not been distinguished from  the 21cmFAST2 model by using either the power spectrum or genus method from the KS test if 1$\sigma$ thermal noise is included.   We also use the $\chi^2$ test and probability to exceed (PTE) to distinguish different models using the 21 cm power spectrum and the 2d genus curve. The $\chi^2$ test is defined as
\begin{equation}
\chi^2=\sum_{i=1}^{N_{\rm bin}}\frac{(y_{i,1}-y_{i,2})^2}{|y_{i,2}|}
\end{equation}
where $y_{i,1}$ and $y_{i,2}$ are the distribution in $i$th bin for model 1 and 2, respectively. $N_{\rm bin}$ is the number of bins and $N_{\rm bin}=31$ for the 2d genus curve and $N_{\rm bin}=20$ for the angular power spectrum. Given an input $\chi^2$ and the number of degrees of freedom $\nu^{\prime}$,  the PTE can be calculated by

\begin{equation}
PTE=\frac{1}{2^{\nu^\prime/2}\Gamma(\nu^\prime/2)}\int_{\chi^2}^{\infty}t^{\nu^\prime/2-1}e^{-t/2}{\rm d}t     
\end{equation}   
A small value of PTE indicates that two models are different. In Table~\ref{tab:prob}, we also give the values of $\chi^2/\nu^\prime$ and PTE for both the power spectrum and genus without and with $1\sigma$ noise. Here, we assume that there are two different parameters when we use the $\chi^2$/PTE test to compare each pair of these simulations. Obviously, the $\chi^2$/PTE tests tell us that the 2d genus curves from different  reionized models can be easily distinguished even 1$\sigma$ thermal noised is added. However, it is difficult to distinguish models HRT sim1 with HRT sim2, HRT sim1 with 21cmFAST2, and HRT2 with 21cmFAST2 from the angular power spectrum method.  It seems that the results from the  $\chi^2$/PTE test are inconsistent with those from the KS test for some comparisons, such as HRT sim2 to 21cmFAST2 by using the 2d genus curves with 1$\sigma$ thermal noise. The reason is that the value of $prob$ from  KS test depends on the deviation between two cumulative distribution function, one discrete data in each bin can not affect the $prob$ significantly, while the value of $\chi^2$ can be vey large even the difference of two models in one bin is significant. Combined with the KS test and $\chi^2$/PTE test for the signals with and without 1$\sigma$ noise, we know that the 2d genus and angular power spectrum are complementary. Moreover, the shape of the power spectrum are nearly the same for the different reionization models, and the only difference is the amplitude, however, it is difficult to obtain the accurate amplitude of the power spectrum in observations. From this perspective the genus method has its niche when compared with the widely used power spectrum.





\section{ Summary and discussion}
We quantify the 2d topology of the 21cm differential brightness  temperature 
field for two HRT simulations and two semi-analytical models. 
It is shown that the 2d topology of $\delta T_b$ is significantly different 
for different reionization models even $1\sigma_T$ thermal noise is added. For the same simulation,
the 2d  topology at different redshifts reflects the status of reioniztion.  

We show the results for both  Gaussian and  compensated  Gaussian beam filter 
of the telescopes.  It is shown that the brightness
temperature maps filtered with these beam patterns can be used to  discriminate 
different reionization scenarios through the study of 
the 2d genus topology. However, the beam filter is more complicated 
in practice, and we need to consider the special case for different telescope.
 Moreover, the foreground removing is crucial for the detection of the neutral HI 
signals, which is beyond our study in the current paper. Of course, this is our 
first step by using the 2d topology of the 21 cm differential brightness to constrain cosmic 
reionization.  The 2d topology can become a very powerful tool for probing the 
reionization history and hope that the real two dimension topology of neutral 
hydrogen at high redshift can be observed by the future telescopes like SKA.

\begin{table*}
\caption{$prob$ values from KS test and $\chi^2/\nu^\prime (PTE) $  from $\chi^2$/PTE test for different reionization  models at $z=9$ with and without 1$\sigma$ thermal noise.}\label{tab:prob}
\begin{center}
\begin{tabular}{lclllllll}\hline

&$prob$  & $prob$  &$\chi^2/\nu^\prime (PTE)$ &$\chi^2/\nu^\prime (PTE)$ \\
& (power spectrum) & (genus) & (power spectrum)& (genus)\\       
\hline\hline
HRT sim1: HRT sim2&0.13 &  0.36  & 0.36(0.99)           & 33.98 (0)        \\
HRT sim1: 21cmFAST1&8.16E-3&0.36    &5.73(0)    &42.22  (0)       \\
HRT sim1: 21cmFAST2&0.28&8.08E-4     &0.89(0.60)          &89.12 (0) \\
HRT sim2: 21cmFAST1&2.57E-3&0.78    &6.48(0)           &77.83 (0)   \\
HRT sim2: 21cmFAST2&0.77&5.65E-3     &0.18(1.00)          &75.48 (0)    \\
21cmFAST1:21cmFAST2&2.57E-3&2.21E-3 & 1.50  (0.08)     &35.39 (0)    \\
\hline\hline
HRT sim1: HRT sim2  $(+1\sigma)$  &0.13 &       0.36  &0.36(0.99)         &137.22(0)     \\
HRT sim1: 21cmFAST1  $(+1\sigma)$ &8.16E-3&     0.36    &5.78(0)     &42.98(0)     \\
HRT sim1: 21cmFAST2  $(+1\sigma)$  &0.28&         0.36   &0.89(0.60)       &184.34(0)  \\
HRT sim2: 21cmFAST1  $(+1\sigma)$ &2.57E-3       &6.21E-2                 & 6.42(0)       &321.39(0)  \\
HRT sim2: 21cmFAST2  $(+1\sigma)$ &0.77&0.99         &           0.18(1.00)        &18.03(0) \\
21cmFAST1:21cmFAST2  $(+1\sigma)$   &2.57E-3& 6.21E-2     &1.50(0.08)       &1168(0)      \\

\hline\hline

\end{tabular}
\end{center}

\end{table*}

\section*{Acknowledgments}
We thank the referee for comments and suggestions that improved the paper.
This  work has started during  YGW's visit to KIAS 2012,  and he would
like to express  his gratitude for KIAS.  We thank Hy Trac and Renyue Cen for providing us the HRT simulation data. We also thank Xin Wang and Bin Yue for many helpful discussions.
This work is supported by the Ministry of Science and Technology 863 project grant 2012AA121701.
YGW acknowledges the 973 Program 2014CB845700, and the NSFC grant 11390372. YDX is supported by NSFC grant No. 11303034.
XLC acknowledges the support  of the 973 program
(No.2007CB815401,2010CB833004), the CAS Knowledge Innovation
Program  (Grant No. KJCX3-SYW-N2), and the NSFC grant 10503010. XLC is also
supported by the NSFC Distinguished Young Scholar Grant No.10525314.



\bibliography{genus_2d}
\bibliographystyle{apj}

\end{document}